\documentclass[
reprint,
superscriptaddress,
preprintnumbers,
nobibnotes,
amsmath,amssymb,
aps,
floatfix,
]{revtex4-1}

\usepackage{bibunits}
\defaultbibliography{Hemozoin}
\defaultbibliographystyle{apsrev4-1}
\usepackage{graphicx}
\usepackage{dcolumn}
\usepackage{bm}
\usepackage{floatrow}
\usepackage{hyperref}
\usepackage{xcolor}
\usepackage{textcomp}
\usepackage{amsmath}
\hypersetup{
    colorlinks,
    linkcolor={red!50!black},
    citecolor={blue!50!black},
    urlcolor={blue!80!black}
}

\newcommand{\unm}{Center for High Technology Materials and Department of Physics and Astronomy, University of New Mexico, Albuquerque, 87106 NM, USA}

\begin{document}

\title{Diamond magnetic microscopy of malarial hemozoin nanocrystals}

\author{Ilja Fescenko}
\email{iliafes@gmail.com}
\affiliation{\unm}

\author{Abdelghani Laraoui}
\affiliation{\unm}

\author{Janis Smits}
\affiliation{\unm}
\affiliation{Laser Center of the University of Latvia, Riga, LV-1586, Latvia}

\author{Nazanin Mosavian}
\affiliation{\unm}

\author{Pauli Kehayias}
\affiliation{\unm}
\affiliation{Department of Physics, Harvard University, Cambridge, 02138 MA, USA}

\author{Jong Seto}
\affiliation{Department of Bioengineering and Therapeutic Sciences, School of Medicine, University of California-San Francisco, San Francisco, 94158 CA, USA}

\author{Lykourgos Bougas}
\affiliation{Johannes Guttenberg University, 55128 Mainz, Germany}

\author{Andrey Jarmola}
\affiliation{Department of Physics, University of California, Berkeley, 94720 CA, USA}
\affiliation{ODMR Technologies Inc., El Cerrito, 94530 CA, USA}

\author{Victor M. Acosta}
\email{vmacosta@unm.edu}
\affiliation{\unm}

\date{\today}

\begin{abstract}
Magnetic microscopy of malarial hemozoin nanocrystals was performed using optically detected magnetic resonance imaging of near-surface diamond nitrogen-vacancy centers. Hemozoin crystals were extracted from \textit{Plasmodium-falciparum}-infected human blood cells and studied alongside synthetic hemozoin crystals. The stray magnetic fields produced by individual crystals were imaged at room temperature as a function of applied field up to 350 mT. More than 100 nanocrystals were analyzed, revealing the distribution of their magnetic properties. Most crystals ($96\%$) exhibit a linear dependence of stray field magnitude on applied field, confirming hemozoin's paramagnetic nature. A volume magnetic susceptibility $\chi=3.4\times10^{-4}$ is inferred using a magnetostatic model informed by correlated scanning electron microscopy measurements of crystal dimensions. A small fraction of nanoparticles (4/82 for \textit{Plasmodium}-produced and 1/41 for synthetic) exhibit a saturation behavior consistent with superparamagnetism. Translation of this platform to the study of living malaria-infected cells may shed new light on hemozoin formation dynamics and their interaction with antimalarial drugs.
\end{abstract}

\maketitle

\begin{bibunit}
\section{\label{sec:Introduction}Introduction}

Magnetic field sensors based on diamond nitrogen-vacancy (NV) centers have emerged as a powerful platform for detecting nanomagnetism in biological samples~\cite{Schirhagl2014,Rondin2014}.  With this technique, magnetic fields from magnetotactic bacteria~\cite{LeSage2013}, ferritin proteins~\cite{Ziem2013}, magnetically labeled cancer cells~\cite{Glenn2015}, and neuronal currents~\cite{Barry2016} have been detected with a remarkable combination of spatial resolution and sensitivity. Diamond magnetic microscopy has even been able to resolve magnetic fields produced by individual nanoparticles exhibiting ferromagnetism and superparamagnetism~\cite{Gould2014,Tetienne2016,Davis2016}. However observation of individual paramagnetic nanoparticles at ambient temperature has remained a challenge, owing to their weaker magnetic signatures.

Of particular interest are paramagnetic hemozoin biocrystals 
that nucleate inside several blood-feeding organisms~\cite{Coronado2014}, including the \textit{Plasmodium} species responsible for the malaria disease. Malarial parasites feed on their host's hemoglobin for essential amino acids, while decomposing the iron complexes into highly toxic, free radicals. These radicals are subsequently bound into chemically inert elongated crystals ($50\mbox{--}1500$ nm in size) called hemozoin~\cite{Coronado2014,Sullivan2002,Hempelmann2007}. 
Hemozoin crystals are a biomarker for malaria disease, and a large effort has been devoted to developing diagnostic platforms based on their detection~\cite{Coronado2014,Cho2012}.  
Hemozoin detection is also used in pharmacological studies of malaria~\cite{Rebelo2013}, since some antimalarial drugs work by altering hemozoin formation~\cite{Hempelmann2007, Sullivan2002}.

\begin{figure*}[ht]
    \centering
\includegraphics[width=.49\textwidth]{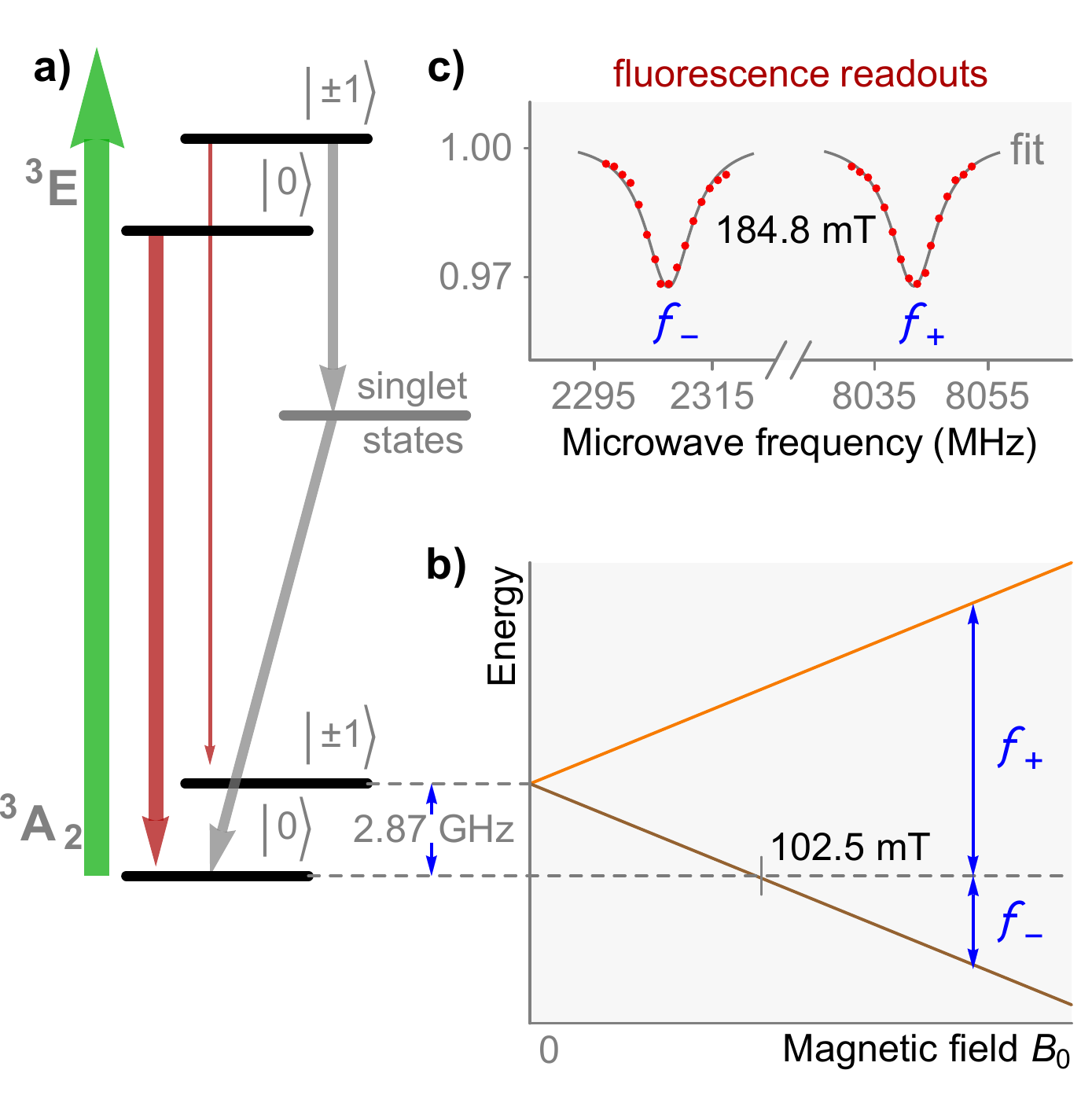}\hfill
\includegraphics[width=.49\textwidth]{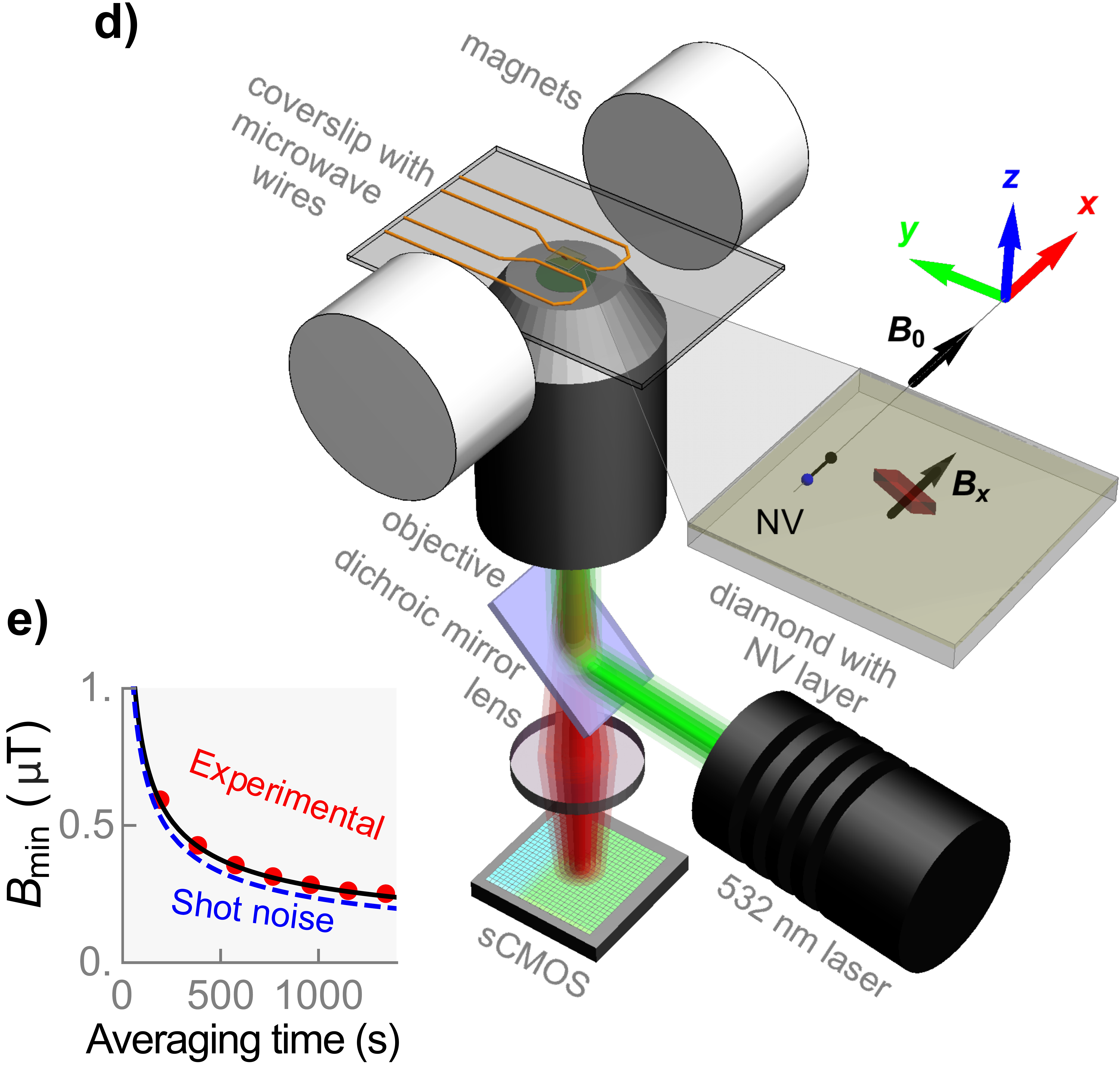}\hfill
       \caption{\label{fig:setup}
\textbf{Diamond magnetic microscopy.}  
		a) NV energy level diagram, depicting magnetic sublevels ($|0\rangle$, $|\pm1\rangle$), optical excitation (green arrow), fluorescence (red arrows), and nonradiative (gray arrows) pathways. Gray arrows show spin-selective intersystem crossing leading to polarization into the \textbar0\big \rangle~ground-state sublevel.  
       b) Energy splitting of the ground-state sublevels in an external magnetic field applied along the N-V axis. Blue arrows indicate the $f_{\pm}$ microwave transitions.  
       c) Example of an optically detected magnetic resonance spectrum. From the separation between peaks, the projection of the local magnetic field along the NV axis is inferred.
       d) Epifluorescence ODMR microscope used for magnetic imaging. Dry hemozoin crystals are placed on top of a diamond substrate with a $\sim0.2$~\textmu m top layer doped with NV centers.  
       e) Experimental and photoelectron-shot-noise-limited detection threshold for $(65~{\rm nm})^2$ detection pixels plotted versus averaging time. The experimental data is fit to the function $B_{\rm min}=\alpha/\sqrt{t}$, with $\alpha=8.4\pm 0.1$~\textmu T$\cdot {\rm s}^{1/2}$.}
\end{figure*}

Hemozoin crystals have characteristic optical properties including birefringence \cite{Lawrence1986}, linear dichroism \cite{Frita2011}, and nonlinear dielectric susceptibility \cite{Belisle2008} that allow for their detection without the use of extrinsic labels. They are also paramagnetic due to the presence of unpaired electrons in their Fe\textsuperscript{3+} centers, meaning they are magnetized only in the presence of an external magnetic field. Direct detection of hemozoin's magnetization is intriguing because, unlike light, magnetic fields are not attenuated by biological tissue.

Several methods of studying hemozoin magnetic signatures have been demonstrated including magneto-optical rotation~\cite{Butykai2013,Orban2016,McBirney2018}, nuclear magnetic resonance relaxometry~\cite{Peng2014,Gossuin2017}, electron paramagnetic resonance ~\cite{Sienkiewicz2006}, and magnetic separation ~\cite{Kim2010}. Direct magnetic detection of hemozoin ensembles has been performed by bulk magnetometry~\cite{Hackett2009,Bremard1992,Bohle1998}. However, detection of stray magnetic fields produced by single hemozoin nanocrystals has not yet been reported, likely due to stringent requirements on sensitivity and spatial resolution. 

We hypothesized that diamond magnetic microscopy~\cite{LeSage2013,Glenn2015,Davis2016,Pham2011,Tetiennee2017} possesses sufficient sensitivity and spatial resolution to image the stray magnetic field produced by individual hemozoin nanocrystals. To test this, we performed room temperature optically detected magnetic resonance (ODMR) imaging of an NV-doped diamond substrate in contact with either ``natural'' (\textit{Plasmodium}-produced) or synthetic hemozoin nanocrystals. Spatially resolved maps of the magnetic fields produced by individual hemozoin nanocrystals were obtained and used to characterize the distribution of their paramagnetic properties. With appropriate modifications, this detection strategy may be used to study the formation dynamics of hemozoin crystals in living malaria-infected cells.

\section{\label{sec:dec}Detection principle}

NV centers are spin-1 defects in the diamond lattice. The energy levels and optical excitation/emission pathways of the NV center are shown in Fig.~\ref{fig:setup}(a).  
For a magnetic field applied along the N-V symmetry axis, the $|0\rangle\leftrightarrow|\pm1\rangle$ spin transition frequencies are  
$f_{\pm}=D\pm\gamma_{NV}B_{\parallel}$, where $D=2.87~{\rm GHz}$ is the zero-field splitting and $\gamma_{NV}=28~{\rm GHz/T}$ is the NV gyromagnetic ratio, Fig.~\ref{fig:setup}(b). The principle of NV magnetometry is to measure these transition frequencies precisely using ODMR techniques~\cite{Schirhagl2014,Rondin2014}. 
When $532~{\rm nm}$ light continuously excites NV centers, the ground state population is optically polarized into $|0\rangle$ via a nonradiative, spin-selective decay pathway involving intermediate singlet states.
Due to the same spin-selective decay mechanism, NV centers excited from $|0\rangle$ emit fluorescence (collected at $650\mbox{--}800$ nm) at a higher rate than those originating from $|\pm1\rangle$. Application of a transverse microwave magnetic field mixes the spin populations, resulting in a dip in fluorescence when the microwave frequency matches the spin transition frequencies, $f_{\pm}$, Fig.~\ref{fig:setup}(c). Monitoring NV fluorescence as the microwave frequency is swept across the resonances reveals $f_{\pm}$ and thus $B_{\parallel}$.

The minimum detectable magnetic field of a Lorentzian ODMR resonance is fundamentally limited by photoelectron shot noise as~\cite{Rondin2014,Acosta2009,Kleinsasser2016}:
\begin{equation}
B_{\rm min}\simeq\frac{4}{3\sqrt{3}}\frac{\Gamma}
{\gamma_{NV}C\sqrt{I_0~t}}\,,
 \label{eq:sen}
\end{equation}
where $\Gamma$ is the full-width-half-maximum (FWHM) linewidth, $C$ is the contrast  (the fractional difference in ODMR signal on/off resonance), $I_0$ is the photoelectron detection rate, and $t$ is the measurement time. With typical experimental values ($\Gamma=12$~MHz, $C=0.02$, $I_0=5\times10^6~{\rm e^{\mbox{-}}/ s}$) for a $(65~{\rm nm})^2$ detection pixel, Eq. \eqref{eq:sen} predicts $B_{\rm min}\simeq7.4$~\textmu T for $t=1~{\rm s}$. Experimentally, we observe a detection threshold that is within $15\%$ of the photoelectron shot noise limit, Fig.~\ref{fig:setup}(e). The experimentally determined $B_{\rm min}$ is calculated as the standard deviation of $(65~{\rm nm})^2$ pixels in magnetic images lacking visible features. Scaling the experimental detection threshold to a $(390~{\rm nm})^2$ pixel (approximately the diffraction-limited resolution), gives $B_{\rm min}\simeq1.4$~\textmu T for $t=1~{\rm s}$. This detection threshold is a factor of 6 lower than for $(65~{\rm nm})^2$ pixels because $I_0$ increases by a factor of 36, Eq. \eqref{eq:sen}.

This detection threshold is sufficient to image the \textmu T fields from hemozoin nanocrystals. In our experimental geometry, Fig. \ref{fig:setup}(d), hemozoin crystals are magnetized along $x$ and the $B_x$ component of the magnetic field is detected by diamond magnetic microscopy. In the point-dipole approximation~\cite{Dolgovskiy2016}, which is valid for crystals with dimensions smaller than the $\sim390~{\rm nm}$ spatial resolution of our microscope, the magnetic field produced from a single crystal (volume, $V$) is:
\begin{equation}
B_x(x,y,z)=\frac{\chi V B_0}{4\pi} \frac{2x^2-y^2-z^2}{(x^2+y^2+z^2)^{5/2}},
 \label{eq:field}
\end{equation} 
where $B_0$ is the applied magnetic field and $\chi$ is the volume magnetic susceptibility. While $\chi$ depends on crystal orientation and purity, values of $3.2\mbox{--}4.6\times10^{-4}$ have been reported in the literature~\cite{Coronado2014,Butykai2013}.
Taking conservative values: $\chi=3.2\times10^{-4}$, $V=({\rm 100~nm})^3$ crystal volume, and $z=200~{\rm nm}$ NV sensing depth, the field produced by a hemozoin nanocrystal in a $B_0=350$~mT applied field has a minimum of $B_x=-1.1$~\textmu T at $x=y=0$ and maxima of $B_x=0.2$~\textmu T at $y=0,x=\pm245~{\rm nm}$. 

To more accurately describe the instrument response, magnetostatic modeling is employed to calculate the $B_x(x,y,z)$ field produced by elongated crystals, with dimensions taken from scanning electron microscopy (SEM) images. We then integrate $z$ over the NV layer distribution, which is assumed to be uniform from $20\mbox{--}200~{\rm nm}$ below the diamond surface (Sec.~\ref{sec:exp}). Finally, we account for the effect of optical diffraction and image drift by convolution with a 2D Gaussian kernel (``blur'') with $0.5$~\textmu m FWHM. 
A discussion of model parameters can be found in Sec.~\ref{sec:unc}.

\begin{figure}[!htb]
   \begin{center}
	\includegraphics[width=\columnwidth]{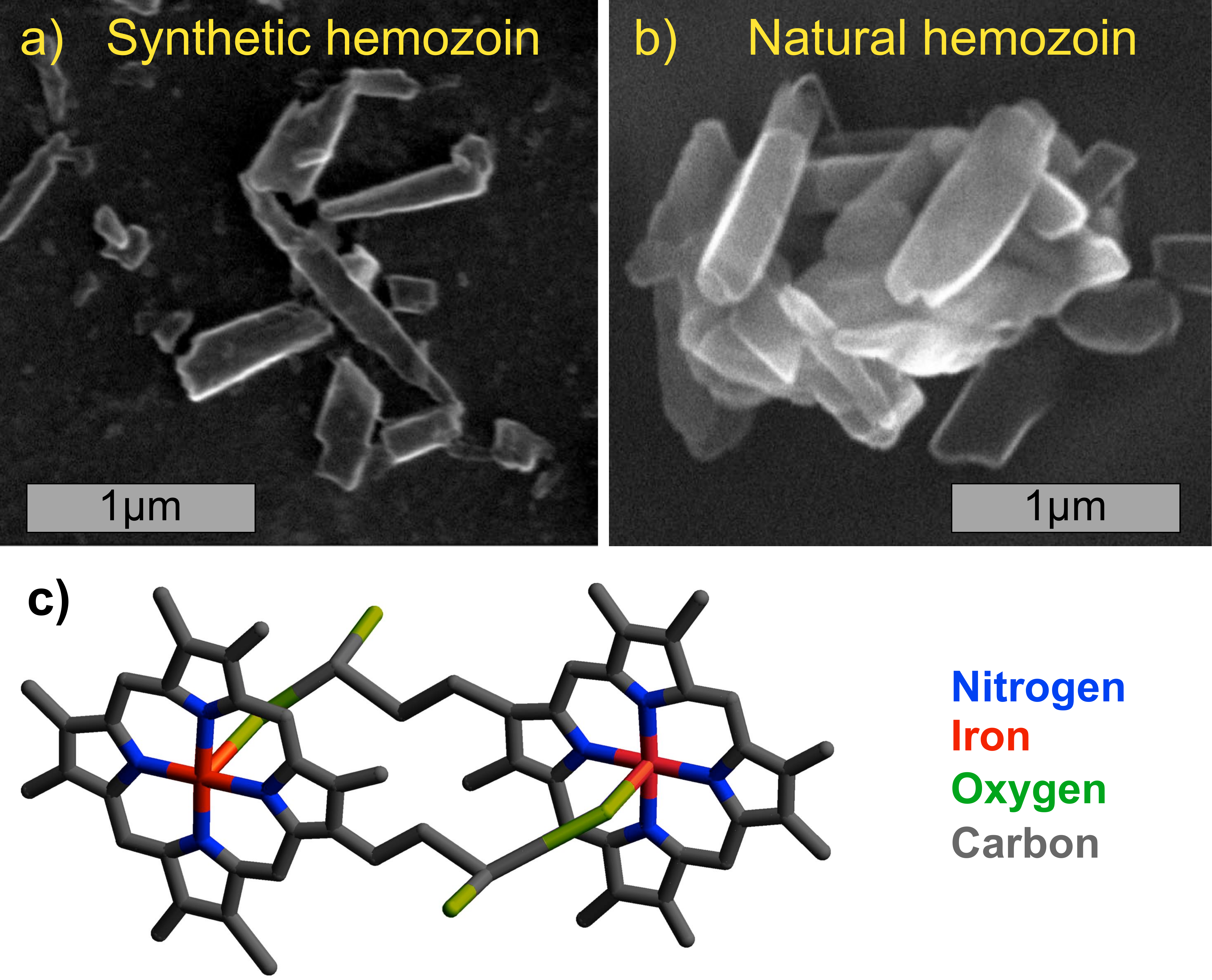}
  \end{center}
    \caption{\label{fig:hems}
\textbf{Hemozoin nanocrystals.} a) SEM image of synthetic hemozoin 
nanocrystals on a diamond substrate. 
b) SEM image of natural hemozoin nanocrystals extracted from human red blood cells co-cultured with \textit{Plasmodium falciparum}, on a diamond substrate.
c) Molecular structure of a hematin dimer, which can form hemozoin crystals when joined together by hydrogen bonds. The molecular structure is expected to be the same for natural and synthetic hemozoin crystals.} 
\end{figure}

\begin{figure*}[ht]
    \centering
	\includegraphics[width=\columnwidth]{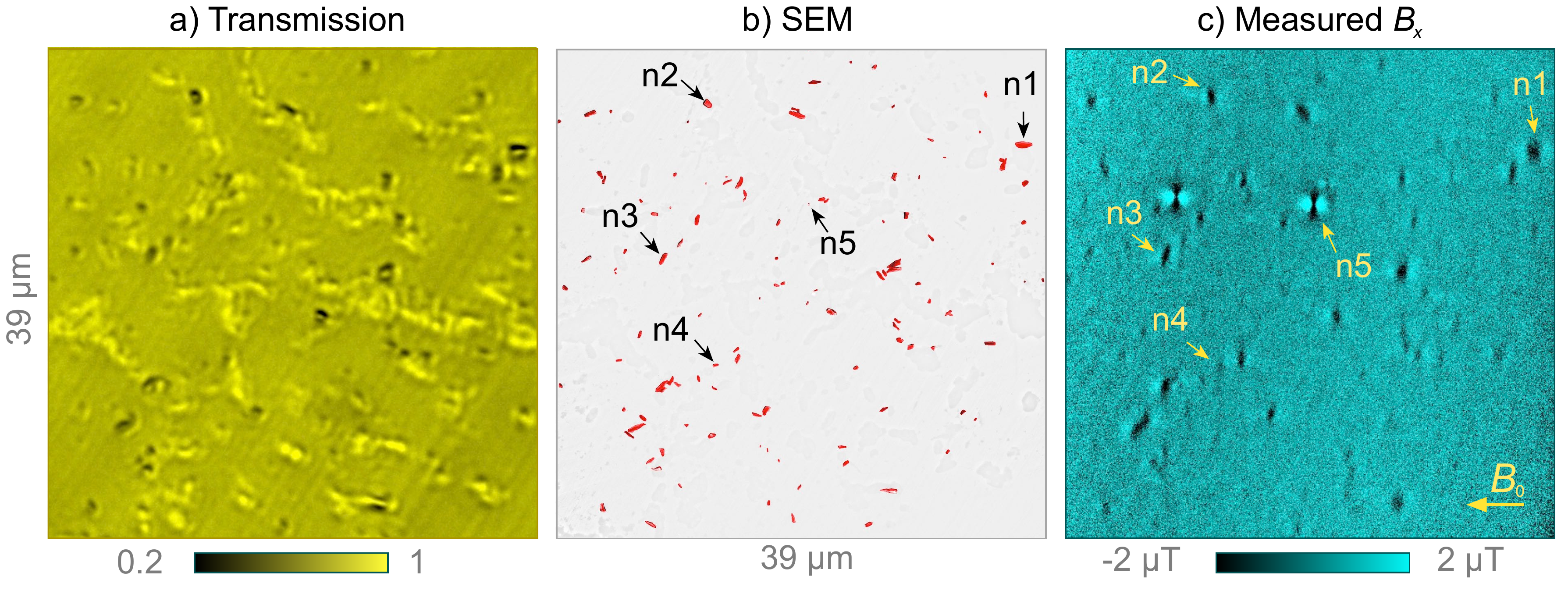}
    \caption{\label{fig:wide}
\textbf{Natural hemozoin.} a) Brightfield transmission image of natural hemozoin crystals dispersed on a diamond substrate. The scale bar is relative transmission. b) SEM image of the same nanocrystals. To enhance hemozoin visibility, the SEM image was modified by a segmentation procedure, depicted in Fig.~\ref{fig:sems}. c) Diamond magnetic microscopy image for applied field $B_0=350~{\rm mT}$. Nanocrystals labeled n1-n5 are studied in detail in Fig. \ref{fig:zoom}(a-e).}
\end{figure*}

\section{\label{sec:Setup}Experimental methods}

Figure \ref{fig:setup}(d) depicts the epifluorescence ODMR microscope used for diamond magnetic microscopy. 
The diamond substrate is a $[110]$-polished, $2\times2\times0.08~{\rm mm}^3$ Type Ib diamond substrate grown by high-pressure-high-temperature synthesis. The substrate was implanted with $^4{\rm He}^+$ ions \cite{Kleinsasser2016} at three different energies (5, 15, and 33 keV) to produce a roughly uniform distribution of vacancies in a $\sim200~{\rm nm}$ near-surface layer, Sec.~\ref{sec:exp}. After implantation, the diamond was annealed in vacuum~\cite{Kehayias2017} at $800^{\circ}~{\rm C}$ (4 hr) and $1100^{\circ}~{\rm C}$  (2 hr) to produce a near-surface layer of NV centers with a density $\sim 10~{\rm ppm}$. Microwaves are delivered by copper loops printed on a contacting glass coverslip. A magnetic field, $\vec{B}_0$, produced by a pair of permanent magnets, points along one of the in-plane NV axes. A linearly polarized 532 nm laser beam (0.2 W) excites the NV centers over a \(\sim\)(40~{\rm\textmu m})$^2$ area, and their fluorescence is imaged onto a sCMOS camera. A detailed description of the apparatus can be found in Sec.~\ref{sec:exp}. 

Magnetic field maps are obtained by performing ODMR imaging of the near-surface NV layer (see Sec.~\ref{sec:proc}). Fluorescence images ($600\times 600$ pixels$^2$, $39\times39$~\textmu m$^2$ field of view, 3~ms exposure time) of the NV layer are recorded at 16 different microwave frequencies around each of the NV ODMR resonances (32 frequencies total). The sequence is repeated and the image set is integrated for several minutes to improve signal-to-noise ratio. The fluorescence intensity vs. microwave frequency data for each pixel are fit by Lorentzian functions
to determine the ODMR central frequencies, $f_{\pm}$. The magnetic field projection along the NV axis is then calculated as $B_{\parallel}\equiv B_0+B_x=(f_{+}-f_{-})/(2\gamma_{NV})$. Determining $B_{\parallel}$ in this way eliminates common-mode shifts in $f_{\pm}$ due to temperature-dependent changes in zero field splitting \textit{D}~\cite{AcostaPRL2010} and variations in longitudinal strain~\cite{Fang2013,Bauch2018}. The external field, $B_0$, is subtracted from the image, revealing a map of the stray magnetic fields, $B_x$, produced by the magnetized nanocrystals. 

Figure \ref{fig:hems} shows SEM images of the synthetic and natural hemozoin nanocrystals studied here. The synthetically-produced hemozoin crystals (InvivoGen\texttrademark, tlrl-hz) have an elongated shape and variable dimensions ($50\mbox{--}1500~{\rm nm})$. The natural hemozoin nanocrystals, extracted from human red blood cells co-cultured with \textit{Plasmodium falciparum}~\cite{Kim2010}, have similar elongated shapes with slightly larger average dimensions. For diamond magnetic microscopy, hemozoin samples were diluted in water and dropcasted on the diamond surface to yield a relatively non-aggregated surface density with approximately 0.04 nanocrystals/\textmu m$^2$.%

\section{\label{sec:results}Results}
Figure~\ref{fig:wide}(a) shows a brightfield transmission image of natural hemozoin crystals dispersed on the diamond sensor's surface. Most of the bright features in the image come from host cell residue; they do not appear in magnetic images. Results from another region without residue are reported in Sec.~\ref{sec:cleannat}. 

Figure~\ref{fig:wide}(b) displays a modified SEM image of the same region as in Fig.~\ref{fig:wide}(a). We used an image segmentation procedure (Sec.~\ref{sec:exp}) to more clearly visualize the hemozoin crystals. Figure~\ref{fig:wide}(c) shows the corresponding magnetic field map obtained by diamond magnetic microscopy. Of a total of 120 features identified as potential hemozoin nanocrystals in the SEM image, 82 exhibit magnetic features resolved by our technique. Surprisingly, the two magnetic features with the largest magnitude correspond to crystals with dimensions $\lesssim200~{\rm nm}$ in the SEM image. Such anomalously bright features were observed consistently, but infrequently (less then $5\%$ of all magnetic features), in both natural and synthetic hemozoin samples. We tentatively attribute them to superparamagnetism~\cite{Inyushin2016}, for reasons discussed below. 

\begin{figure*}[ht]
    \centering
	\includegraphics[width=\columnwidth]{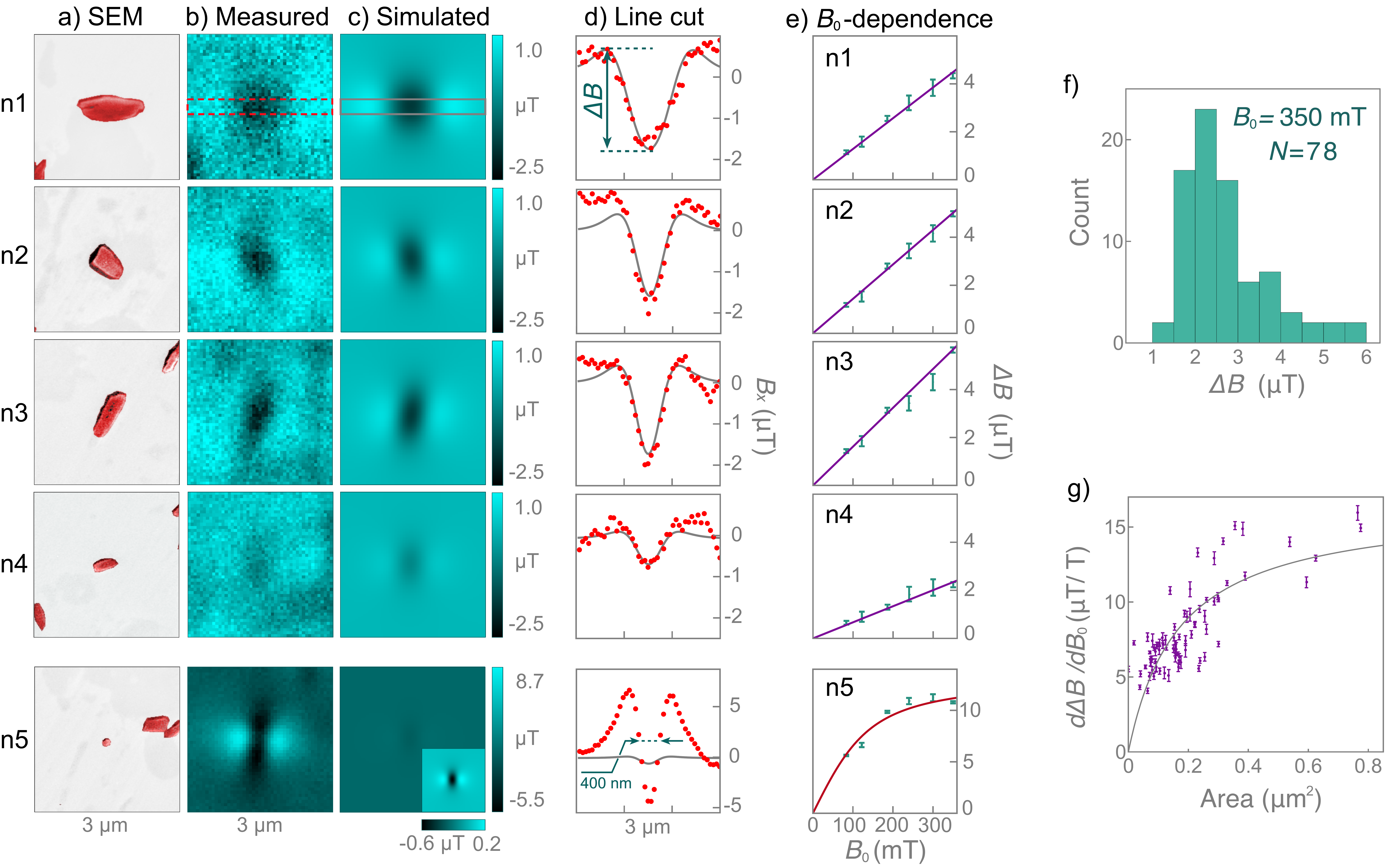}
   
    \caption{\label{fig:zoom}
\textbf{Magnetic imaging of individual natural hemozoin nanocrystals.} 
a) SEM images of hemozoin nanocrystals n1-n5, labeled on Fig.~\ref{fig:wide}.
b) Corresponding diamond magnetic microscopy images ($B_0=186$~mT) for each crystal. 
c) Simulated magnetic images ($\chi=3.4\times10^{-4}$, $B_0=186$~mT) using nanocrystal dimensions inferred from (a). Only the crystal at the center of each SEM image was included in the model. 
d) Line cuts of each nanocrystal (red: measured, gray: simulated), from which the field pattern amplitude $\mathit\Delta B$ was inferred. A minimum feature width of $\sim400$ nm FWHM is observed for n5, close to the optical diffraction limited resolution of our microscope. 
e) $\mathit\Delta B(B_0)$ for each crystal. Solid lines are weighted fits to a line (n1-n4) or Langevin function (n5). All fits are constrained to intercept the origin, $\mathit\Delta B(0)=0$ (zero coercivity assumption). 
f) Histogram of $\mathit\Delta B~(B_0=350~{\rm mT})$ of the 78 crystals exhibiting linear, paramagnetic behavior (including n1-n4). The four crystals exhibiting superparamagnetic behavior (including n5) were excluded from the analysis. 
g) Fitted slopes, $d\mathit\Delta B/dB_0$, as a function of crystal area, $A$, as determined from SEM images. The data are fit with an empirical saturation function, $d\mathit\Delta B/dB_0=S_{\rm max}/(1+A_{\rm sat}/A$), with $S_{\rm max}=16.5\pm 1.4$~\textmu T/T and $A_{\rm sat}=0.17\pm 0.03$~\textmu m$^2$.}
\end{figure*}

Five example crystals are labeled n1-n5 on the images in Fig.~\ref{fig:wide}(b-c), including one of the aforementioned strongly magnetized nanocrystals (n5). Figure~\ref{fig:zoom}(a) shows SEM images of each example crystal; n1-n4 have a typical size and shape for these crystals, whereas n5 is barely visible, with dimensions $<200~{\rm nm}$. Figure~\ref{fig:zoom}(b) shows the corresponding diamond magnetic microscopy images taken at $B_0=186~{\rm mT}$. Each crystal exhibits a different field pattern characteristic of its unique size, shape, and orientation.  

Figure~\ref{fig:zoom}(c) shows the expected magnetic field patterns of each nanocrystal, calculated using the procedure described in Sec.~\ref{sec:dec}. Each nanocrystal was modeled as a 3D ellipsoid with uniform susceptibility and dimensions inferred from the corresponding SEM images. The height of the crystals was assumed to be 200 nm. For n1-n4, the model produces similar field patterns to those observed experimentally. The pattern amplitude is best described using a volume magnetization $M=50~{\rm A/m}$, which corresponds to a volume susceptibility of $\chi=\mu_0 M/B_0=3.4\times10^{-4}$, where $\mu_0=4\pi\times10^{-7}~{\rm m\cdot T/A}$ is the vacuum permeability. This value of $\chi$ is well within the range of literature values for hemozoin~\cite{Coronado2014,Butykai2013} and is in good agreement for most crystals, despite a wide variation in their magnetic patterns. This demonstrates that crystal size, shape, and orientation are the primary factors in determining the magnetic pattern behavior. Factors contributing to a $\lesssim25\%$ uncertainty in $\chi$ are discussed in Sec. \ref{sec:unc}.

For crystal n5, however, the model does a poor job of describing the magnetic behavior. It predicts a field pattern amplitude more than an order of magnitude lower than the observed value, suggesting n5 is not paramagnetic hemozoin. It is difficult to estimate this particle's magnetization, since it may arise from a small inclusion or adjacent particle not resolved in the SEM image.

Line cuts of the magnetic field patterns for n1-n5 are shown in Fig.~\ref{fig:zoom}(d). The line cuts are obtained by averaging $B_x$ values over 6 rows (390~nm) in a band along the magnetic feature, as indicated in Fig. \ref{fig:zoom}(b-c). The line cuts were used to calculate the magnetic pattern amplitude $\mathit\Delta B$, determined from the difference in extreme $B_x$ values in the line cuts. The uncertainty in $\mathit\Delta B$ is $\sim\pm0.2$~\textmu T, based on the scatter in the $B_x$ values in the line cuts in a magnetic-feature-free region. The amplitudes for n1-n4 vary due to size/shape, but all fall in the range $\mathit\Delta B=1.3\mbox{--}2.9$~\textmu T. However the amplitude for n5 is much larger, $\mathit\Delta B=11.0\pm 0.2$~\textmu T. The width of this feature is $\sim400$ nm FWHM, close to our microscope's diffraction-limited spatial resolution, providing further evidence that it comes from a point-like particle.

Magnetic images of all nanocrystals shown in Fig.~\ref{fig:wide} were obtained for six values of external field: $B_0$= 83, 122, 186, 240, 300, and 350 mT. From these images, we identified 82 individual magnetized crystals and calculated $\mathit\Delta B$ as a function of $B_0$ for each. The $\mathit\Delta B(B_0)$ curves for all 82 crystals are provided in Fig.~\ref{fig:fitnat}. Fig~\ref{fig:zoom}(e) shows the curves of nanocrystals n1-n5. A linear dependence is found for n1-n4 and is characteristic of paramagnetic response. In total, 78 out of 82 natural hemozoin crystals show a similar linear behavior, with a slope in the range $d\mathit\Delta B/dB_0=4\mbox{--}16$~\textmu T/T. 

The remaining four nanoparticles from the natural hemozoin sample exhibit a saturation-like behavior, as seen for n5. These curves were fit with a Langevin function of the form $\mathit\Delta B(B_0) = a[\coth(b B_0)-1/(b B_0)]$, where $a$ and $b$ are fit parameters. For n5, $a=13.3\pm 0.7$~\textmu T and $b=0.017\pm 0.002~{\rm mT}^{-1}$. This model is commonly used to describe superparamagnetic nanoparticles in the limit that the thermal energy exceeds the magnetic anisotropy energy~\cite{Langevin}. The large magnetization and saturation behavior in these small nanoparticles is consistent with recent reports of superparamagnetism in hemozoin samples ~\cite{Inyushin2016,Khmelinskii2017}. However we point out that only a small fraction ($\lesssim5\%$) of nanoparticles exhibit such behavior. This may explain why superparamagnetic behavior was not observed in previous ensemble studies with unrefined samples~\cite{Gossuin2017,Hackett2009,Butykai2013}. The atomic structure of these outlier nanoparticles remains a topic for future work.

A histogram of $\mathit\Delta B$ for all 78 paramagnetic nanocrystals at $B_0=350$ mT is shown in Fig.~\ref{fig:zoom}(f). The distribution is characterized by a mean amplitude of $2.9$~\textmu T, a median of $2.6$~\textmu T, and a standard deviation of $1.6$~\textmu T. Figure~\ref{fig:zoom}(f) plots the best-fit slopes, $d\mathit\Delta B/dB_0$, as a function of crystal area, as determined from SEM images. The slopes increase in a roughly monotonically fashion before saturating when the crystal dimensions exceed the spatial resolution of the microscope. This behavior is expected for crystals with uniform susceptibility. For crystal volumes much smaller than the sensing voxel, $V<<V_{\rm sense}\approx0.4\times0.4\times0.2~{\rm \mu m}^3$, Eq.~\eqref{eq:field} predicts $d\mathit\Delta B/dB_0\propto \chi V$, while for $V>>V_{\rm sens}$ we expect $d\mathit\Delta B/dB_0\propto \chi$ independent of crystal dimensions. 

\begin{figure}[!t]
    \begin{center}
	\includegraphics[width=\columnwidth]{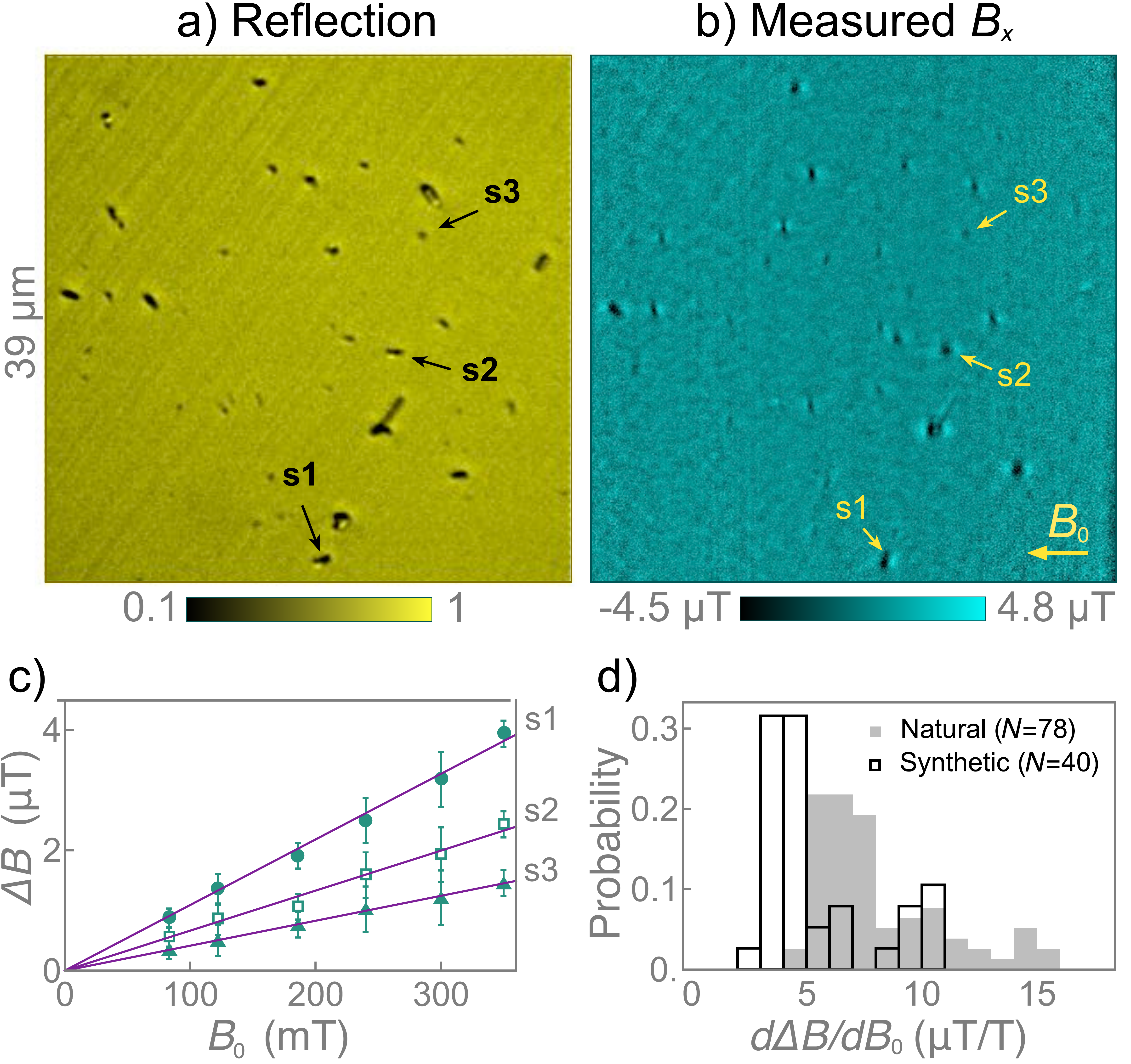}
    \end{center}
    \caption{\label{fig:hist}
\textbf{Synthetic hemozoin.} a) Confocal reflection image (excitation 405 nm) of synthetic hemozoin nanocrystals on a diamond substrate. In total, 46 dark features are identified as likely hemozoin nanocrystals. b) Corresponding diamond magnetic microscopy image at $B_0=350~{\rm mT}$. We find that 41 out of 46 possible crystals exhibit an observable magnetic feature. c) $\Delta B(B_0)$ curves for three synthetic hemozoin crystals labeled in b). Solid lines are weighted linear fits. d) Histogram of $d\mathit\Delta B/dB_0$ for natural and synthetic hemozoin crystals.}
\end{figure} 

We also studied the magnetic properties of synthetic hemozoin crystals manufactured by InvivoGen\texttrademark. These nanocrystals are commonly used as a model for natural hemozoin, owing to their ease of procurement and nearly identical crystal morphology and chemical structure \cite{Jaramillo2009,Oliveira2005}. Previous studies reported similarities in the ensemble magnetic properties of synthetic and \textit{Plasmodium falciparum} extracted hemozoin \cite{Orban2014}; however, to our knowledge, this is the first study comparing the distribution of their magnetic properties at the single nanocrystal level. 

Figure~\ref{fig:hist}(a) shows a confocal reflection image of dried synthetic hemozoin crystals dispersed on the diamond substrate. Out of a total of 46 dark features identified as potential hemozoin, 41 produce magnetic features in the diamond magnetic microcopy images, Fig.~\ref{fig:hist}(b). As with natural hemozoin, $\mathit{\Delta} B(B_0)$ curves were generated by monitoring these features in magnetic images taken at six different applied fields. The $\mathit{\Delta} B(B_0)$ curves for all 41 nanocrystals are displayed in Fig.~\ref{fig:fitsyn}. One nanocrystal exhibited a Langevin saturation, suggesting superparamagnetic behavior, while the remaining 40 crystals exhibited linear dependence. Fig.~\ref{fig:hist}(c) shows the $\mathit{\Delta} B(B_0)$ curves for three example nanocrystals exhibiting linear behavior, labeled s1-s3 in Fig.~\ref{fig:hist}(b). Magnetostatic modeling [Fig.~\ref{fig:zoomsyn}(d)] of these three nanocrystals show good agreement with the experimental images using $\chi=3.4\times10^{-4}$. This is the same value as found for natural hemozoin [Fig.~\ref{fig:zoom}(c-d)], indicating the natural and synthetic crystals have similar magnetic properties.

Figure~\ref{fig:hist}(d) shows histograms of the fitted slopes, $d\mathit{\Delta} B/dB_0$, for all 40 paramagnetic synthetic crystals and all 78 paramagnetic natural crystals. The synethtic nanocrystals have a slightly smaller slope (mean=$5.5$, median=$4.3$~\textmu T/T) compared to the natural hemozoin (mean=$8.1$, median=$7.2$~\textmu T/T), while the standard deviation is similar ($6.2$ and $8.0$~\textmu T/T, respectively). This is likely due to a small difference in size distribution, evident in the SEM images in Fig.~\ref{fig:sems}.

\section{\label{sec:summary}Outlook and conclusion}

Our results demonstrate the capability of diamond magnetic microscopy to simultaneously measure the magnetic properties of numerous individual biocrystals. Whereas bulk measurements yield ensemble-average properties, diamond magnetic microscopy can measure the distribution of nanocrystal susceptibilities, extract information about each crystal's dimensions and orientation, and differentiate paramagnetic hemozoin from superparamagnetic nanoparticles. This brings up the intriguing possibility of using this platform to monitor the formation dynamics of individual hemozoin nanocrystals in living cells without the use of extrinsic contrast agents. 

To assess the feasibility of performing magnetic imaging of hemozoin inside cells, we estimate a typical hemozoin crystal volume is $V=0.04$~\textmu m$^3$, based on the median area found from SEM images ($0.2$~\textmu m$^2$) and an assumed thickness of $0.2$~\textmu m. We conservatively assume the hemozoin crystal has formed in a digestive vacuole at the far periphery of a \textit{Plasmodium}-infected red blood cell (typical thickness: 2~\textmu m~\cite{Cho2012}) adhered to the diamond surface. Such a crystal, with susceptibility $\chi=3.4\times10^{-4}$, under an applied field $B_0=350$~mT, would produce a magnetic pattern amplitude at the diamond surface ($z=2$~\textmu m) of $\mathit{\Delta} B=
0.057$~\textmu T, Eq. \eqref{eq:field}. This field shift would be detectable with a signal:noise ratio of 1 after $\sim10~{\rm min}$ of averaging using the present diamond sensor with $(390~{\rm nm})^2$ detection pixels [see Sec.~\ref{sec:dec}]. However the 200 nm NV layer currently used is optimized for negligible diamond-hemozoin standoff. If the standoff is $\sim2~\rm$ \textmu m, an NV layer of 1~\textmu m would be more optimal, leading to a signal:noise ratio of 1 after $\sim2~{\rm min}$. If a higher magnetic field is used (already realized for nanoscale NV magnetometry~\cite{Stepanov2015,Nabeel2015}), a signal:noise ratio of 1 should be obtained for $B_0=3~{\rm T}$ after $\sim1.6~{\rm s}$ of signal averaging. 

This sensitivity would be sufficient to monitor individual hemozoin nanocrystals on minute-to-hour timescales. It would enable monitoring of the position and orientation of crystals throughout the parasite life cycle under different conditions, including the influence of antimalarial drugs. If the platform is simplified and miniaturized, it may also find application as a label-free malaria diagnostic with a sensitivity rivaling the current staining/microscopy standard \cite{Butykai2013}. Our platform can also be used to study paramagnetic substances other than hemozoin. Figure~\ref{fig:hemin} shows diamond magnetic microscopy images of the pharmacological agent hemin.

In summary, we used diamond magnetic microscopy to characterize the distribution of magnetic properties of synthetic and natural hemozoin samples at the individual nanocrystal level. More than $95\%$ of the nanocrystals exhibited paramagnetic behavior, with magnetic field patterns well described by a magnetostatic model using a volume susceptibility $\chi=3.4\times10^{-4}$. Five out of the 123 nanocrystals studied exhibited anomalously large magnetization that saturated at fields above $\sim0.1~{\rm T}$, suggesting superparamagnetism. Future work is needed to determine the composition and structure of these outlier nanoparticles. With minor improvements in the experimental setup, diamond magnetic microscopy should be capable of imaging hemozoin formation dynamics in living cells, with implications for malaria diagnosis and drug development. 

\begin{acknowledgments}
The authors acknowledge valuable conversations with C. Belthangady, D. Budker, Y. Silani, and J. Damron, as well as sample characterization resources at the UNM Comprehensive Cancer Center Fluorescence Microscopy Shared Resource and the DOE Center for Integrated Nanotechnologies (CINT). This work was funded by NIH (NIGMS) award 1R41GM119925-01, NIH (NIMH) award 1R41MH115884-01, NIH (NIBIB) award 	1R01EB025703-01, and a Beckman Young Investigator award.

\textbf{Competing interests}

L. Bougas, A. Jarmola, and V. M. Acosta are co-inventors on a related patent: US 2018/0203080 A1. A. Jarmola and L. Bougas are co-founders of startup ODMR Technologies and have financial interests in the company. The remaining authors declare no competing financial interests.

\textbf{Author contributions}

L. Bougas, P. Kehayias, and (later) V. M. Acosta conceived of the idea in consultation with A. Jarmola. P. Kehayias performed initial calculations demonstrating feasibility. I. Fescenko, A. Jarmola, and V. M. Acosta designed the experimental plan with input from all authors. J. Smits, P. Kehayias, and I. Fescenko wrote and implemented control and automation software. J. Seto extracted the natural hemozoin samples. I. Fescenko, A. Laraoui, and N. Mosavian built the measurement apparatus, prepared samples, and collected data. I. Fescenko analyzed the data and wrote the initial manuscript draft, under supervision of V. M. Acosta. All authors discussed results and contributed to the writing of the manuscript.
\end{acknowledgments}

\putbib
\end{bibunit}

\widetext
\clearpage

\begin{center}
\textbf{\large Supplemental Information: Diamond magnetic microscopy of malarial hemozoin nanocrystals.}
\end{center}
\setcounter{equation}{0}
\setcounter{section}{0}
\setcounter{figure}{0}
\setcounter{table}{0}
\setcounter{page}{1}
\setcounter{equation}{0}
\setcounter{figure}{0}
\setcounter{table}{0}
\setcounter{page}{1}
\makeatletter
\renewcommand{\thetable}{S\arabic{table}}
\renewcommand{\theequation}{S\arabic{equation}}
\renewcommand{\thefigure}{S\arabic{figure}}
\renewcommand{\thesection}{S\Roman{section}}
\renewcommand{\bibnumfmt}[1]{[S#1]}
\renewcommand{\citenumfont}[1]{S#1}

\begin{bibunit}
\section{\label{sec:hem}Hemozoin}

Hemozoin crystals consist of dimers of \(\alpha\)-hematin molecules each containing an iron center. Figure~\ref{fig:hems} in the main text shows the molecular structure of a dimer, constructed using the Avogadro open-source molecular builder and visualization tool. The central iron of the first \(\alpha\)-hematin is bonded to the oxygen of the carboxylate side-chain of the second \(\alpha\)-hematin. The dimers are joined together by hydrogen bonds to form a triclinic crystal~\cite{Egan2002}. These crystals are alternatively called ``hemozoin'' or ``\(\beta\)-hematin''. Throughout this manuscript, we use the term hemozoin to refer to crystals produced both naturally (through biocrystallization inside living organisms) and synthetically. 

The organisms that produce hemozoin crystals include the various \textit{Plasmodium} species, the parasitic worms \textit{Schistosoma mansoni}, the avian protozoan parasite \textit{Hemoproteus columbae}, and the kissing bug \textit{Rhodnius prolixus}~\cite{Egan2002}. In humans hemozoin plays an important role in the pathology of the \textit{Plasmodium} parasites responsible for malaria infection, a disease which continues to have a devastating effect in tropical and subtropical regions~\cite{Coronado2014}. 

During their life cycle, malarial protozoans feed on host-cell hemoglobin by decomposing it inside digestive vacuoles into amino acids and toxic \(\alpha\)-hematin radicals~\cite{Coronado2014,Cho2012,Sullivan2002,Hempelmann2007}. The iron radicals are then bound, inside the same vacuoles, into inert hemozoin crystals and released into the host blood during erythrocyte disintegration. 

The suppression (or alteration) of hemozoin formation inside \textit{Plasmodium} parasites is considered a key action of widely used antimalarial drugs, like quinine and chloroquine~\cite{Hempelmann2007,Sullivan2002}. Quinine was successfully used for more than 300 years to treat malaria by suppressing hemozoin crystallization~\cite{Hempelmann2007}. Chloroquine was widely used in the 20\textsuperscript{th} century and thought to invoke a similar response in hemozoin formation~\cite{Hempelmann2007,Sullivan2002}. The regular use of these drugs has lead to the selection of resistant forms of \textit{Plasmodium} parasites, which carry genes responsible for preventing the drugs from entering digestive vacuoles ~\cite{Hempelmann2007,Sullivan2002}.  

The motivation for the present study was to establish diamond magnetic microscopy as an effective tool for fast, quantitative, non-invasive characterization of individual hemozoin nanocrystals. Translation of this technique to the study of hemozoin formation in living protozoans could be used to elucidate the interaction between hemozoin formation, the host immune system, and antimalarial drugs. This line of inquiry may even lead to the establishment of a bank of effective antimalarial drugs able to counteract the highly adaptive parasites. 

\section{\label{sec:exp}Experimental details}

\begin{figure*}[!b]
    \centering
\includegraphics[width=.49\textwidth]{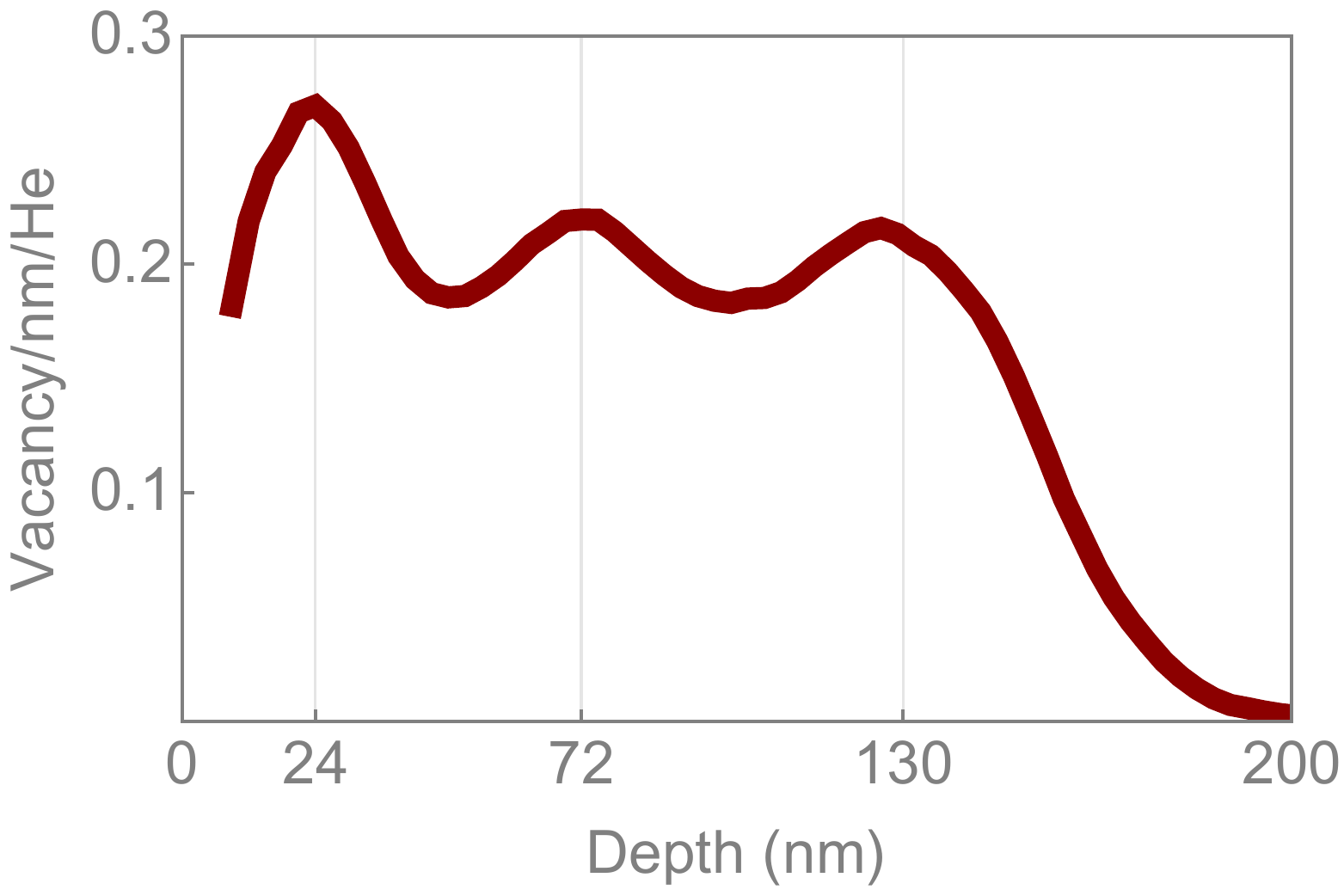}\hfill
\includegraphics[width=.49\textwidth]{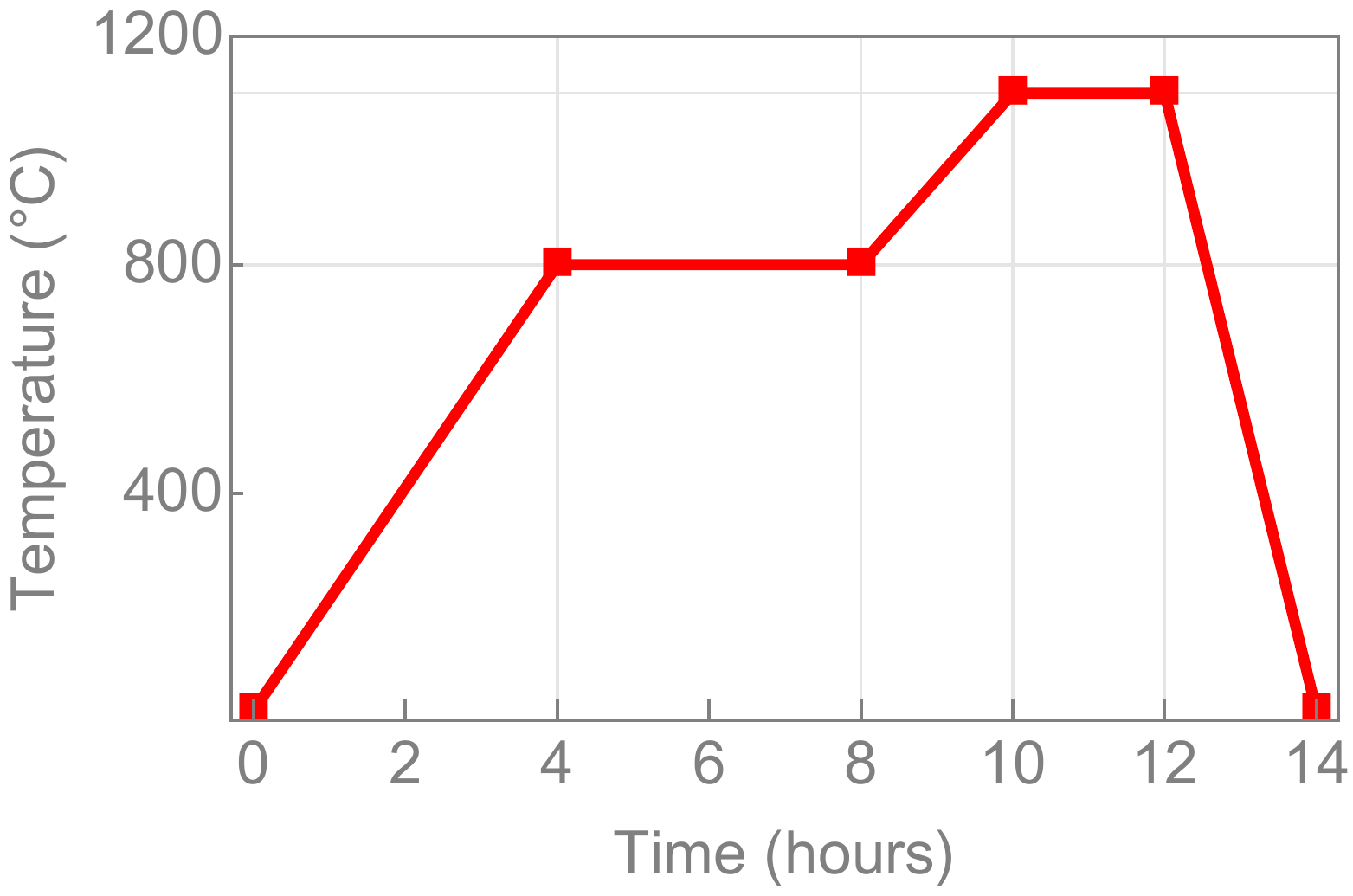}\hfill
       \caption{\label{fig:srim}
\textbf{NV layer fabrication.}  
	a) SRIM vacancy depth profile for implantation conditions discussed in the text. 
       b) Time-temperature graph for the annealing procedure used in this study. 
       }
\end{figure*}

\begin{figure}[!b]
   \begin{center}
	\includegraphics[width=\columnwidth]{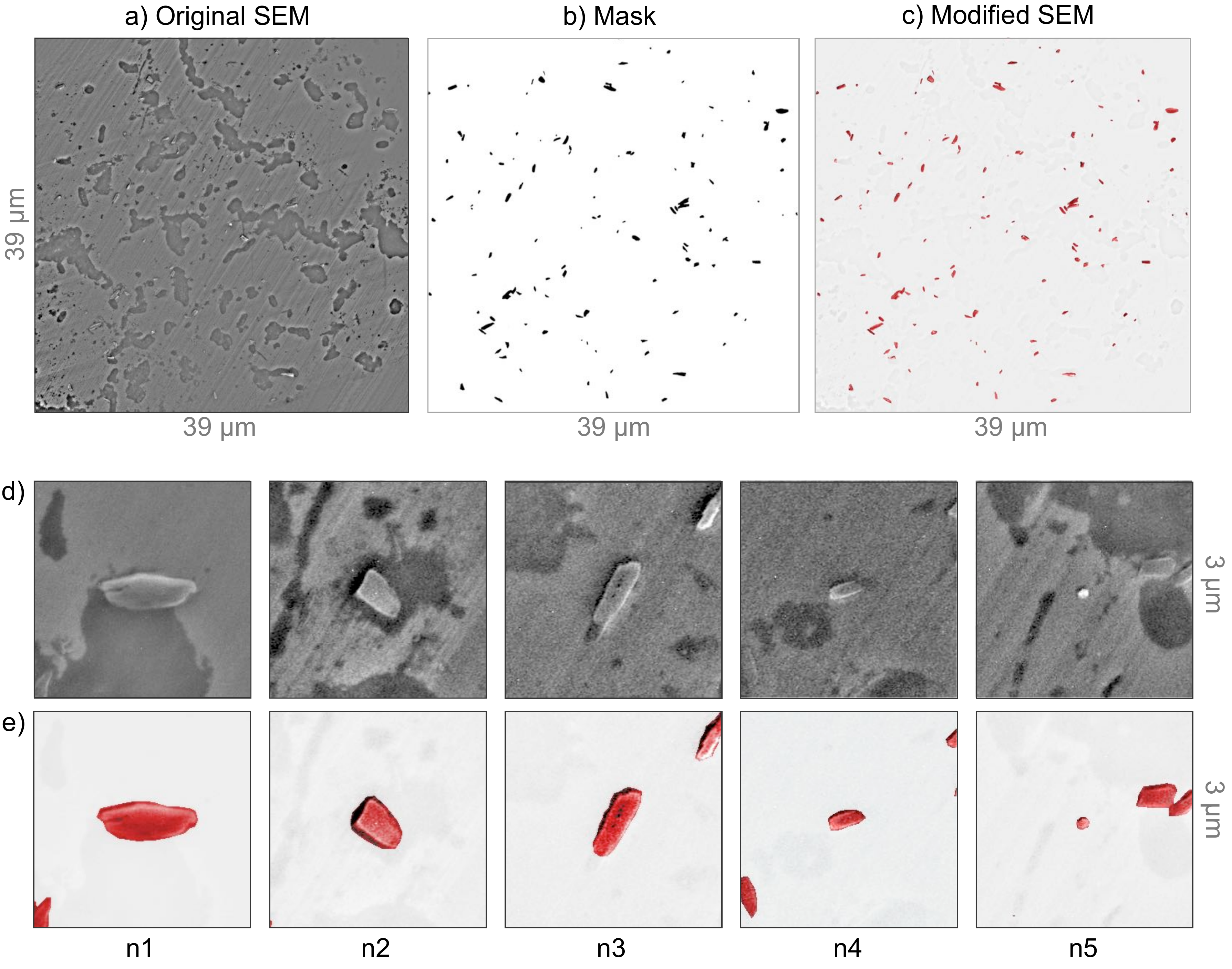}
  \end{center}
    \caption{\label{fig:sems}
\textbf{SEM processing.} a) Unprocessed SEM micrograph of the natural hemozoin sample. b) A mask used to highlight and colorize hemozoin crystals and to attenuate their background. c) Modified SEM as shown in Fig.~\ref{fig:wide}(b) in the main text. (d) Unprocessed and (e) modified SEM images of n1-n5 hemozoin crystals addressed in Fig.~~\ref{fig:zoom}(a-e) }  
\end{figure}

The diamond substrate used in this work is a $[110]$-polished, $2\times2\times0.08~{\rm mm}^3$ Type Ib diamond substrate grown by high-pressure-high-temperature (HPHT) synthesis. The substrate was implanted with $^4{\rm He}^+$ ions \cite{Kleinsasser2016} at three different energies (5, 15, and 33 keV) to produce a roughly uniform distribution of vacancies in a $\sim200~{\rm nm}$ near-surface layer. Vacancy distribution depth profiles were estimated using the Stopping and Range of Ions in Matter (SRIM) Monte-Carlo simulation, Fig.~\ref{fig:srim}(a). The diamond substrates were modeled as a pure carbon layer with 3.52 g/cm$^3$ density and 37.5 eV atom displacement threshold energy. The lattice damage threshold and surface damage threshold were set to 7.35 eV and 7.5 eV, respectively. Note that SRIM simulations do not take into account crystallographic effects such as ion channeling, and therefore could lead to an underestimation of the vacancy layer depth, but are sufficiently accurate for our purposes. Our SRIM simulations predict a $\sim$24 nm modal depth (the depth where vacancy density is greatest) for 5 keV $^4{\rm He}^+$ ion implantation, $\sim$72 nm depth for 15 keV, and $\sim$130 nm depth for 33 keV. We used helium implantation doses of $4\times 10^{-12}$~He/cm$^2$, $2\times 10^{-12}$~He/cm$^2$, and $2\times 10^{-12}$~He/cm$^2$ for, respectively 5, 15, and 33 keV implantation energy. SRIM simulations, Fig.~\ref{fig:srim}(a), indicate this produces a vacancy density of $\sim50~{\rm ppm}$ with $\lesssim30\%$ variation throughout the layer. After implantation, the diamond was annealed in vacuum at $800^{\circ}~{\rm C}$ and $1100^{\circ}~{\rm C}$, Fig.~\ref{fig:srim}(b). This process resulted in a $\sim200~{\rm nm}$ near-surface layer of NV centers with a density $\sim 10~{\rm ppm}$, as determined by fluorescence intensity~\cite{Acosta2009}. The largest source of uncertainty in the final NV layer distribution is the extent of vacancy diffusion during annealing. Previous studies~\cite{Santori2009} predict tens of nanometers of diffusion during annealing of Type 1b HPHT diamonds under similar conditions. We thus expect a slight broadening of the vacancy distribution in Fig.~\ref{fig:srim}(a) to $\sim200~{\rm nm}$ after annealing. 

To magnetize nanocrystals, we used the static magnetic field created by a pair of permanent magnets (Main text, Fig.~\ref{fig:setup}(d). The magnetic field was aligned to the [110]-polished diamond surface such that $\vec{B}_0$ was parallel to one of the two possible in-plane NV symmetry axes. 
Microwave fields were delivered by two copper loops (each addressing a different spin transition) printed on a glass coverslip. The loops were positioned with respect to the diamond to maximize the microwave field components orthogonal to the NV axis. The microwaves were generated by a SRS SG384 frequency generator with two outputs for the first and second harmonics. The frequency of the generator was swept by sending a ramp function to the generator's frequency-modulation (FM) analog input. For $B_0>186$~mT, additional frequency doublers (Minicircuits ZX90-2-24-S+ and ZX90-2-36-S+) were used on both harmonic outputs to access higher microwave frequencies. The microwaves for each transition were amplified separately via MiniCircuits amplifiers: ZHL-16W-43-S+ (for MW frequencies below 4 GHz), ZVE-3W-83+ (for 4-8 GHz), and ZVE-3W-183+ (for $>8$ GHz). The output of the amplifiers was subsequently routed to the coverslip-printed loops. 

To excite NV fluorescence, $\sim0.2~{\rm W}$ of $532~{\rm nm}$ light from a Lighthouse Photonics Sprout-G-10W laser was used in a homebuilt inverted epifluorescence microscope [main text, Fig.~\ref{fig:setup}(d)]. We used an NA=1.3 oil-immersion objective, but an air gap between the coverslip and diamond reduced the diffraction-limited spatial resolution to $\sim390~{\rm nm}$. The laser light was focused to the lateral periphery of the objective's back focal plane to produce a $\sim(40$~\textmu m$)^2$ illumination region on the diamond's top surface (the surface in contact with the dry hemozoin). By exciting in this way, the laser beam entered the bottom of the transparent diamond substrate at a steep angle, leading to substantial internal reflection from the top diamond-air interface. This was helpful in maximizing NV excitation intensity, while minimizing the intensity incident on hemozoin samples. We found no evidence of photodamage on hemozoin samples even after studying them under intense, continuous illumination for several days. The microscope was operated on a standard vibration-isolation optical table in a climate-controlled laboratory environment. The fluorescence was spectrally filtered by a dichroic mirror and a Semrock BrightLine® 731/137 nm bandpass filter, focused by a 200 mm focal length achromatic lens (Thorlabs ITL200), and detected by a Hamamatsu Orca v2 sCMOS sensor. By alternating a flip mirror, the fluorescence could also be detected by a Thorlabs APD410A avalanche photo diode connected to a Yokogawa DL9040 oscilloscope for faster tuning and alignment. The same microscope configuration was used to obtain brightfield transmission images by collecting white light transmitted through the analyte and diamond. 

A LabVIEW program was used to control the experiment. A software command triggered the camera to initiate a burst of sixteen 3-ms frames in rolling shutter mode. Just before the first frame, the camera provided a precisely timed TTL pulse which was used to trigger a sequence generated by a National Instruments multifunction Data Acquisition (DAQ) module. The DAQ provided an analog ramp function which was used to sweep the microwave generator frequency via its analog FM modulation input. Sixteen frames were acquired for each microwave sweep. After each sweep, the microwave generator was reconfigured via GPIB command to alternate between first and second harmonics to alternately address $f_+$ or $f_-$ transitions. Camera frame acquisition and real-time image processing was synchronized via the LabVIEW-interface. The image processing procedure is described in detail in Sec.~\ref{sec:proc}.

Synthetic hemozoin crystals were purchased from InvivoGen\texttrademark (tlrl-hz) and natural hemozoin crystals were obtained from human red blood cells co-cultured with \textit{Plasmodium falciparum} parasites. The hemozoin samples were centrifuged several times (5 min at 6000 rpm) in de-ionized water to wash out salts and cell residues. They were subsequently sonicated to minimize aggregation, and finally homogeneously dispersed on the diamond surface by dropcasting the hemozoin suspension and wicking away the water until it had dried.

The shape, size, and orientation of the hemozoin crystals were investigated by SEM (UNM Center for High Technology Materials and DOE Center for Integrated Nanotechnologies) and confocal reflectance microscopy (UNM Comprehensive Cancer Center Fluorescence Microscopy Shared Resource). An SEM image for the synthetic hemozoin sample reported in the main text was not obtained due to the sample's early destruction. However confocal reflectance images (405 nm excitation) were obtained, which offer a spatial resolution of $\sim225~{\rm nm}$.

Figure~\ref{fig:sems}(a) shows the raw SEM image of the natural hemozoin sample studied in the main text. The faint slanted line pattern is due to imperfections in the diamond polishing. Although the raw SEM images were of high enough quality to identify and measure hemozoin crystals, there was some background due to cell residue. We thus post-processed the raw SEM image in order to enhance hemozoin visibility in the main text, Fig.~\ref{fig:wide}(b). The hemozoin crystals were selected using the free selection tool in graphical redactor GIMP, and a mask was created from the selection, Fig.~\ref{fig:sems}(b). 
The mask was used to colorize the selected crystals, while the inverted mask was used to diminish the contrast of the background.   

\begin{figure}[!t]
   \begin{center}
	\includegraphics[width=\columnwidth]{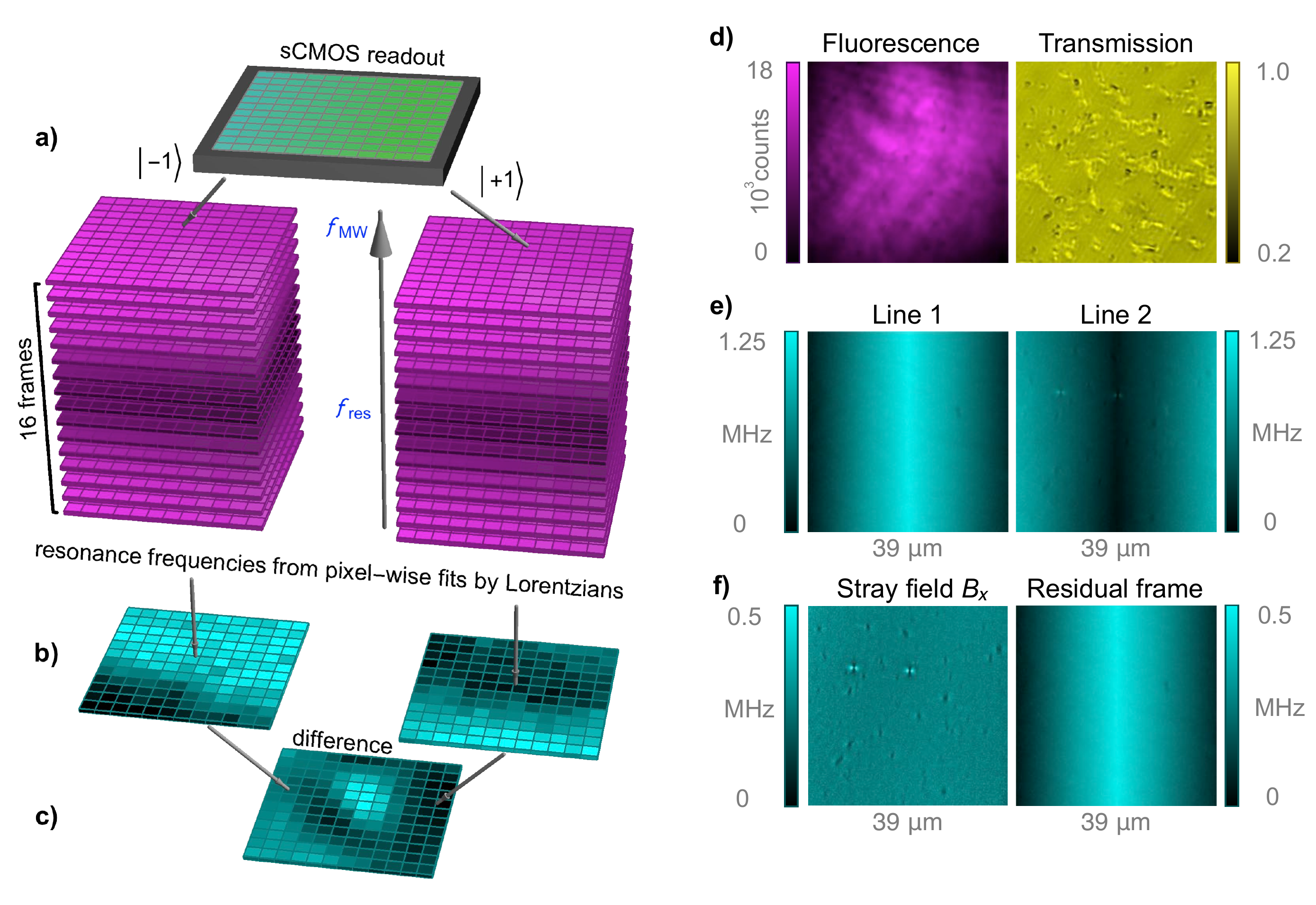}
  \end{center}
    \caption{\label{fig:frame}
\textbf{Data processing.} a) Fluorescence images are acquired at each of 32 different microwave frequencies (16 frequencies for each of the $f_{\pm}$ spin transitions). b) For any given pixel, two 16-point ODMR curves (fluorescence intensity versus microwave frequency) are generated. Each ODMR curve is fit to a Lorentzian function to reveal the ODMR central frequencies for that pixel. This is repeated for each pixel, yielding two maps of ODMR central frequencies corresponding to $f_+$ and $f_-$. c) The magnetic field pattern is obtained by subtracting the images for $f_+$ and $f_-$, dividing by $2\gamma_{NV}$, and subtracting off the applied $B_0$ field. d) An example fluorescence image for a fixed microwave frequency along with a brightfield transmission image of the same field of view. e) Intermediate step showing separate maps for $f_+$ and $f_-$ frequencies. f) The final map, taken at $B_0=350$~mT, is proportional to stray magnetic fields, $\gamma_{NV} B_x=(f_{+}-f_{-})/2-\gamma_{NV} B_0$. This is shown alongside a map of residual non-magnetic shifts, $(f_{+}+f_{-})/2-2D$. Note that $f_{-}$ is defined such that it is negative when $|-1\rangle$ is lower in energy than $|0\rangle$. The horizontally-varying pattern is an artifact of the sCMOS camera's dual-sensor readout. This artifact is present when operating in rolling shutter mode, but is eliminated using our subtraction procedure. }  
\end{figure}

\section{\label{sec:proc}Data processing}

The principle of our image processing procedure is shown in Fig.~\ref{fig:frame}. While sweeping the microwave frequency in a range of $\sim12~{\rm MHz}$ about each ODMR transition, a set of 32 arrays of fluorescence images is acquired from the sCMOS sensor and accumulated for signal averaging. Each of the images are comprised of $600\times600$ pixels. Each pixel corresponds to a $65\times65$ nm$^2$ area in the sample plane, Fig.~\ref{fig:frame}(d). The images are arranged in two 3D image stacks with dimensions $16\times600\times600$, where the first dimension is arranged by microwave frequency. By fitting Lorentzian functions along the microwave-frequency dimension, two $600\times600$ images corresponding to each ODMR central frequency are produced, Fig.~\ref{fig:frame}(b,e). 

A value of the central frequency probed in a particular pixel of the sensor contains information about the local magnetic field, $B_0+B_x$. However other factors, such as strain or temperature~\cite{AcostaPRL2010}, can also produce frequency shifts. Fortunately, the difference of the two ODMR frequencies gives a pure measure of the magnetic field, $f_+-f_-=2\gamma_{NV}(B_0+B_x)$. Subtracting off the contribution from $B_0$ gives the final magnetic images as in Fig.~\ref{fig:frame}(c,f). 

The horizontally varying pattern in Fig.~\ref{fig:frame}(e) is actually varying along columns of the sCMOS sensor (the images are rotated by $90^\circ$ in our setup). This is an artifact of using the sCMOS sensor's rolling shutter mode. The sCMOS camera is comprised of two sensors, and the readout from each sensor starts from the center rows and moves out to the edges of the CMOS chip. The total readout time takes $\sim3~{\rm ms}$, which is approximately the same as the exposure time. This means that rows in the center see a different microwave frequency from the rows at the edges. The difference is approximately the sweep range divided by the number of microwave points, $12~{\rm MHz}/16=0.75~{\rm  MHz}$. To eliminate this artifact, we scanned the microwave frequencies across both ODMR resonances in the same direction for $B_0<102.5~{\rm mT}$ (both $f_{\pm}$ are positive), whereas for $B_0>102.5~{\rm mT}$ ($f_{\pm}$ have opposite signs) we scanned the microwave frequencies in opposite directions. This scanning and subtraction procedure was effective in removing this artifact, Fig.~\ref{fig:frame}(f).

In this work, we used 16 frequencies per ODMR resonance, as we found that near-shot-noise-limited performance could be realized in this manner without introducing artifacts. This lead to a maximum magnetic image acquisition rate of $\sim10~{\rm Hz}$. While we performed Lorentzian fitting offline, if one were to implement this post-processing procedure in real time, the image rate may be further reduced. 
However, in principle, only two frequencies per resonance are necessary, and in this manner the ODMR positions could be determined algebraically without Lorentzian fitting. This would allow for magnetic image refresh rates $\sim100~{\rm Hz}$ enabling video recording of fast dynamics. 
 
\begin{figure}[!b]
    \begin{center}
	\includegraphics[width=0.6 \columnwidth]{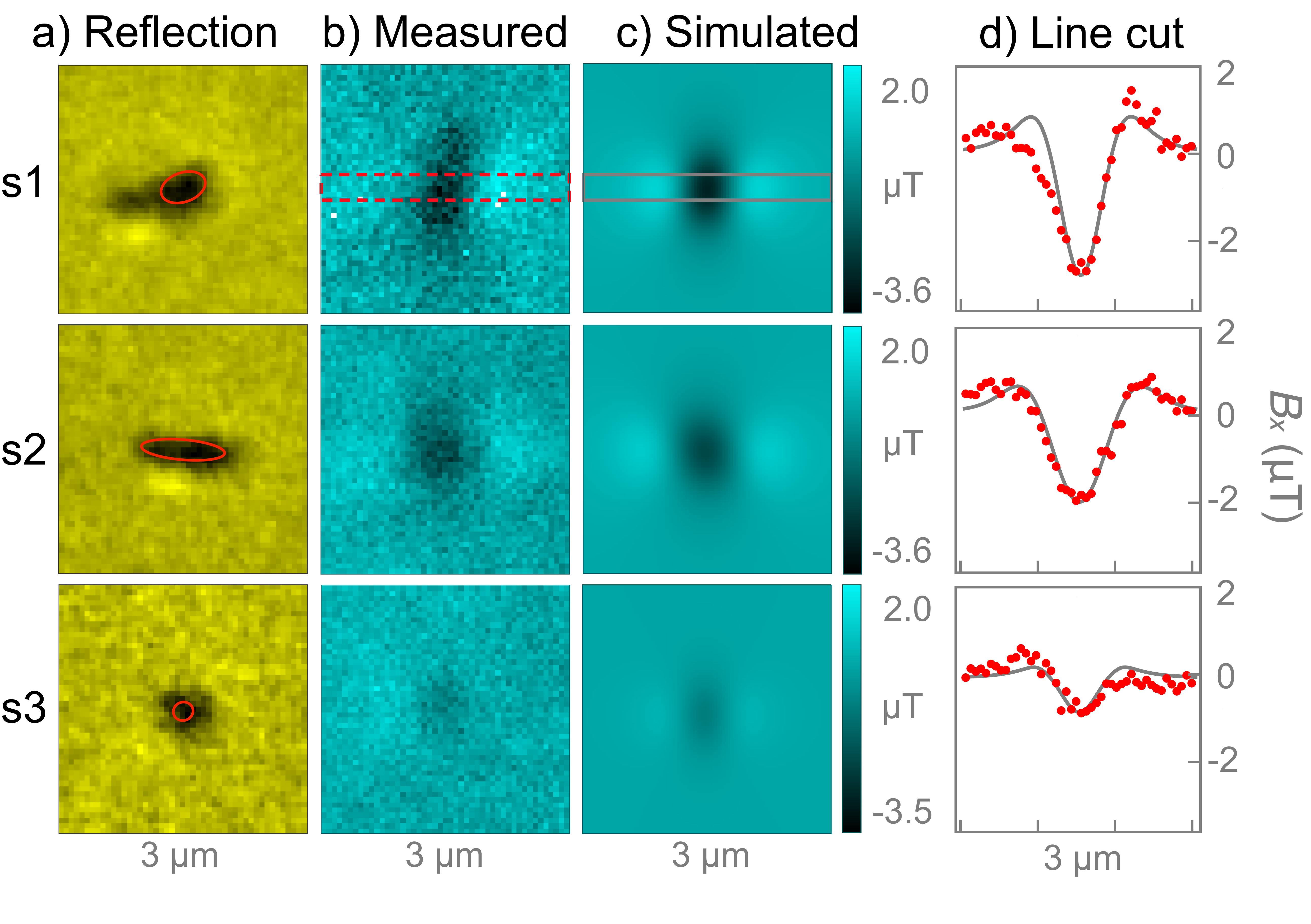}
    \end{center}
    \caption{\label{fig:zoomsyn}
\textbf{Individual synthetic hemozoin crystals.}
a) Confocal reflection images of hemozoin nanocrystals s1-s3, labeled on Fig.~\ref{fig:hist}(a-b) in the main text. The estimated area used for magnetostatic modeling is outlined in red. 
b) Corresponding diamond magnetic microscopy images ($B_0=350$~mT) for each crystal. 
c) Simulated magnetic images ($\chi=3.4\times10^{-4}$, $B_0=350$~mT) using nanocrystal dimensions roughly estimated from (a).  
d) Line cuts of each nanocrystal (red: measured, gray: simulated), from which the field pattern amplitude $\mathit\Delta B$ was inferred. }
\end{figure}

\section{\label{sec:fits}Auxiliary results} 
\subsection{\label{sec:zoomsyn} Magnetostatic modeling of s1-s3}
Three individual crystals, labeled s1-s3 on the images in Fig.~\ref{fig:hist}(b-c, main text), are addressed in detail in Fig.~\ref{fig:zoomsyn}.  Figure~\ref{fig:zoomsyn}(a) shows confocal reflection images of each crystal. Figure~\ref{fig:zoomsyn}(b) shows the corresponding diamond magnetic microscopy images taken at $B_0=186~{\rm mT}$. As with the natural hemozoin sample shown in the main text, each synthetic crystal exhibits a different field pattern characteristic of its unique size, shape, and orientation.  

Figure~\ref{fig:zoomsyn}(c) shows the expected magnetic field patterns of each nanocrystal, calculated using the procedure described in Sec.~\ref{sec:dec} of the main text. Each nanocrystal was modeled as a 3D ellipsoid with uniform susceptibility and dimensions roughly estimated from the corresponding confocal reflection images. The height of the crystals was assumed to be 200 nm. For all crystals, the model produces similar field patterns to those observed experimentally. The pattern amplitude is best described using a susceptibility of $\chi=3.4\times10^{-4}$, which is the same as the value of $\chi$ used for natural hemozoin crystals in the main text.  

Line cuts of the magnetic field patterns for s1-s3 are shown in Fig.~\ref{fig:zoomsyn}(d). The line cuts are obtained by averaging $B_x$ values over 6 rows (390~nm) in a band along the magnetic feature, as indicated in Fig.~\ref{fig:zoomsyn}(b-c). The line cuts were used to calculate the magnetic pattern amplitudes $\mathit\Delta B$ at each value of $B_0$. The results of this analysis are displayed in Fig.~\ref{fig:hist}(c-d) in the main text and Fig. \ref{fig:fitsyn} below.

\subsection{\label{sec:cleannat} Natural hemozoin in a second field of view}
\begin{figure}[!t]
   \begin{center}
	\includegraphics[width=\columnwidth]{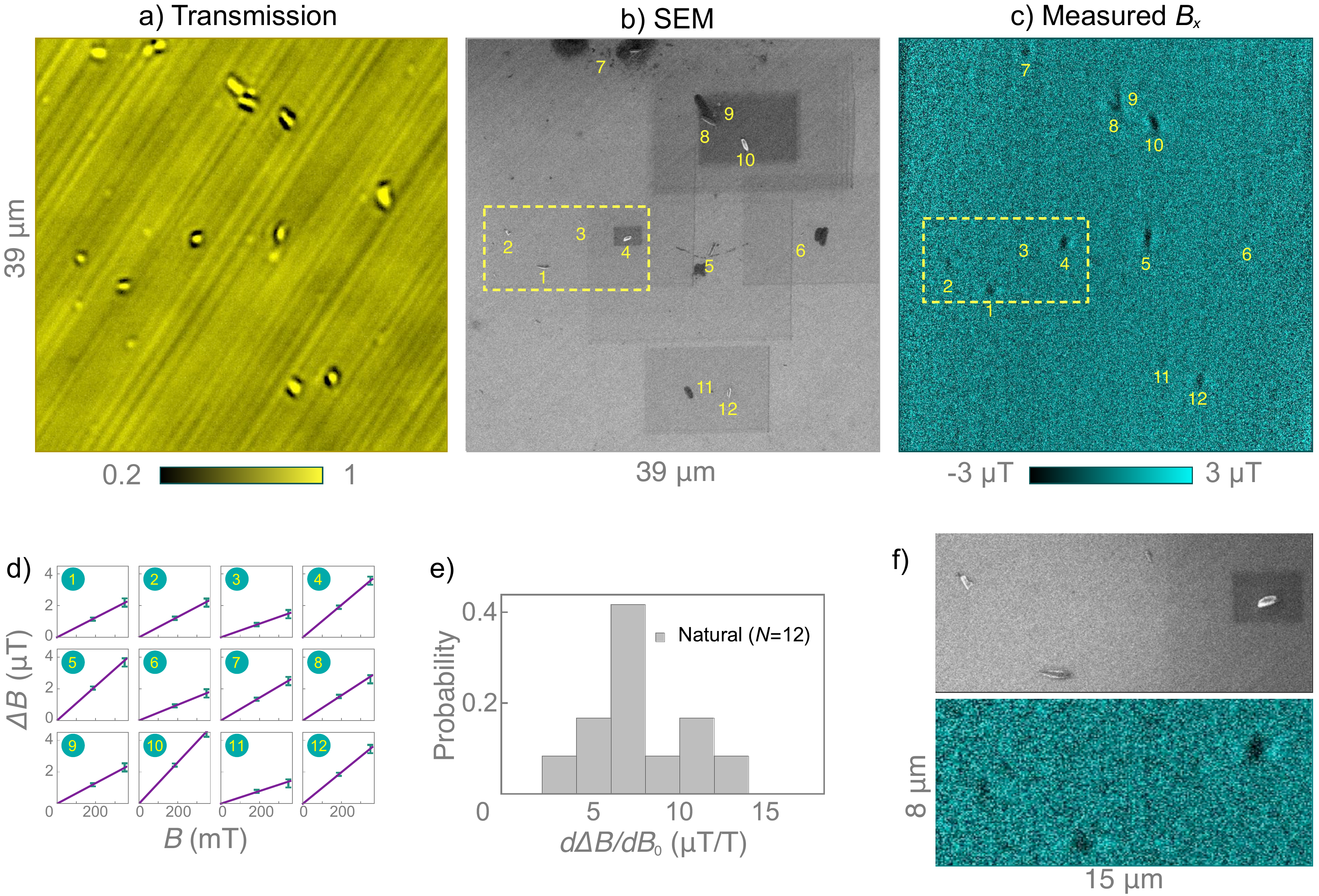}
  \end{center}
    \caption{\label{fig:spot1}
  \textbf{A second region of the natural hemozoin sample.} a) Brightfield transmission image of natural hemozoin crystals dispersed on a diamond substrate. b) SEM image of the same nanocrystals. Dark rectangles are due to overcharging during zooming. c) Corresponding diamond magnetic microscopy image at $B_0=350~{\rm mT}$ showing positions of each nanocrystal.
d) $\mathit\Delta B(B_0)$ curves for all 12 hemozoin crystals. 
e) Histogram of $d\mathit\Delta B/dB_0$ determined from the linear slopes in e). 
f) Magnified SEM and magnetic images of crystals 1-4.}  
\end{figure}

In addition to the hemozoin samples in the main text, we also studied natural hemozoin crystals in a second region of the diamond substrate. Figure~\ref{fig:spot1}(a-c) shows, respectively, brightfield transmission, SEM, and diamond magnetic microscopy images in this second field of view. The hemozoin density is sparser in this region (only 12 crystals are identified) and less cell residue is present. Figure~\ref{fig:spot1}(d) shows the $\mathit\Delta B(B_0)$ curves for each of the 12 crystals, and Fig.~\ref{fig:spot1}(e) shows the histogram of the fitted slopes. The distribution is similar to the one for natural hemozoin found in Fig.~\ref{fig:hist}(d) in the main text. Figure~\ref{fig:spot1}(f) shows a magnified view of the SEM and magnetic images containing four crystals.

\begin{figure}[!htb]
   \begin{center}
	\includegraphics[width=\columnwidth]{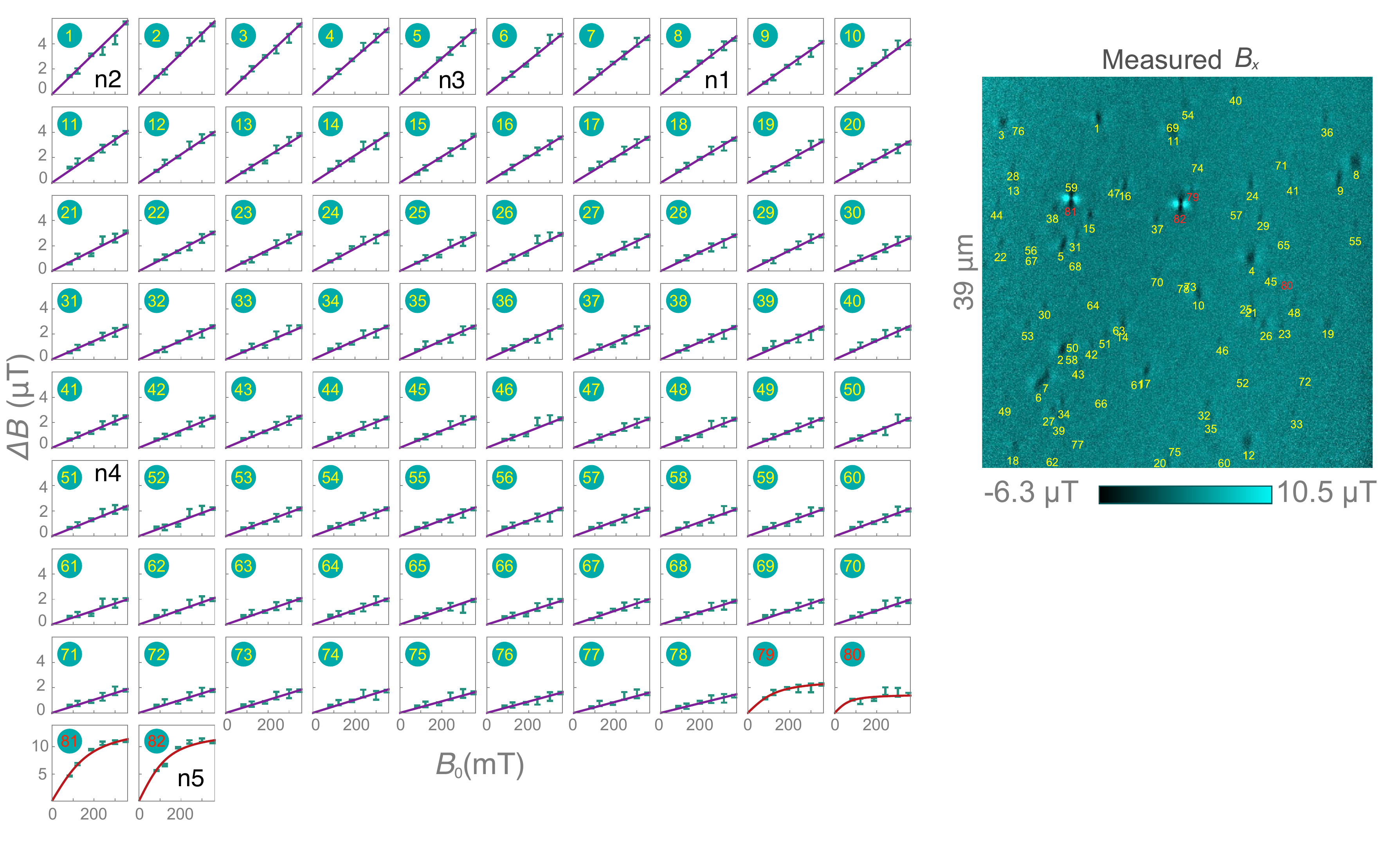}
  \end{center}
    \caption{\label{fig:fitnat}
\textbf{Natural hemozoin: field-dependent magnetizaton.} (left) $\mathit\Delta B(B_0)$ curves for all 82 natural hemozoin crystals reported in the main text. (right) Magnetic image ($B_0=350~{\rm mT}$ showing positions of each nanocrystal. }  
\end{figure}

\begin{figure}[!htb]
   \begin{center}
	\includegraphics[width=\columnwidth]{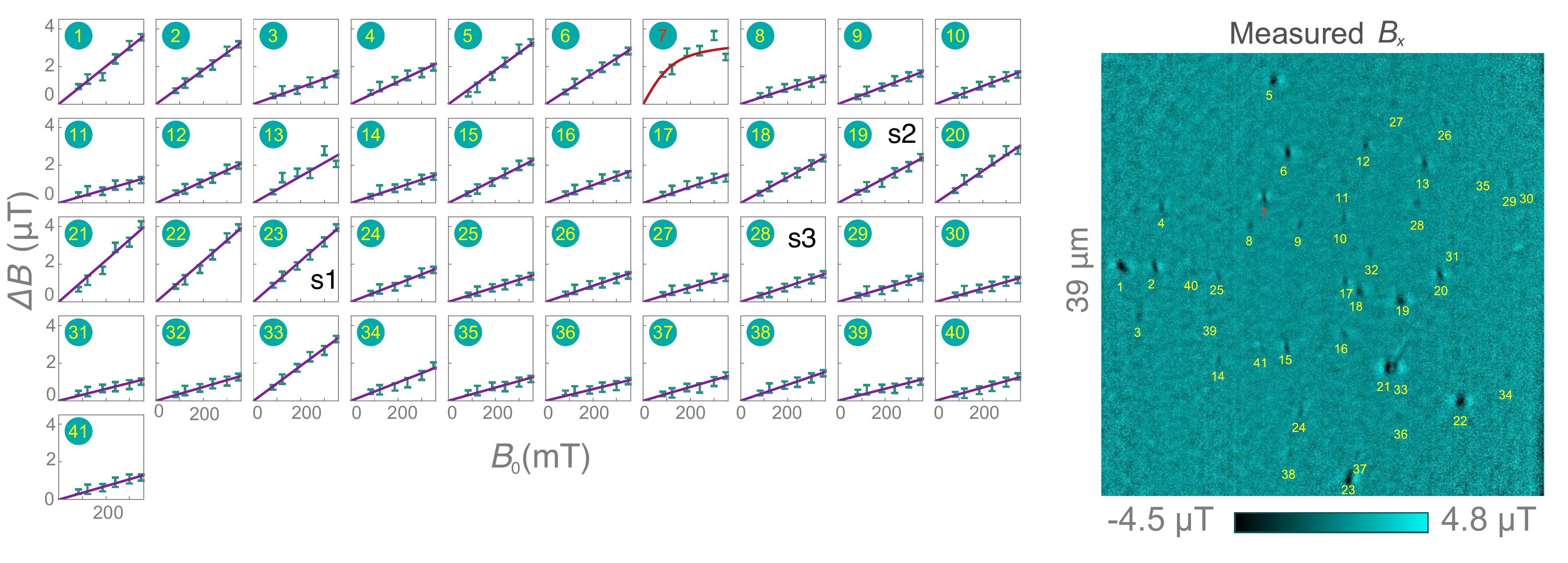}
  \end{center}
    \caption{\label{fig:fitsyn}
\textbf{Synthetic hemozoin: field-dependent magnetizaton.} (left) $\mathit\Delta B(B_0)$ curves for all 41 synthetic hemozoin crystals reported in the main text. (right) Magnetic image ($B_0=350~{\rm mT}$ showing positions of each nanocrystal. }  
\end{figure}
\subsection{\label{sec:completecurves} Complete field-dependent magnetization data set for samples studied in main text}
Figure~\ref{fig:fitnat} shows $\mathit\Delta B(B_0)$ curves for all 82 observed magnetic features in the natural hemozoin sample. The paramagnetic crystals are labeled with yellow numbers, while particles exhibiting superparamagnetic-like behavior are labeled in red. The corresponding magnetic features are labeled by numbers on the widefield magnetic image (taken at $B_0=350~{\rm mT}$). 

Figure~\ref{fig:fitsyn} shows $\mathit\Delta B(B_0)$ curves for the 41 magnetic features observed in the synthetic hemozoin sample reported in the main text. The particle exhibiting superparamagnetic behavior is labeled in red. The corresponding magnetic features are labeled by numbers on the widefield magnetic image (taken at $B_0=350~{\rm mT}$).  

\begin{figure}[!t]
    \begin{center}
	\includegraphics[width=0.6 \columnwidth]{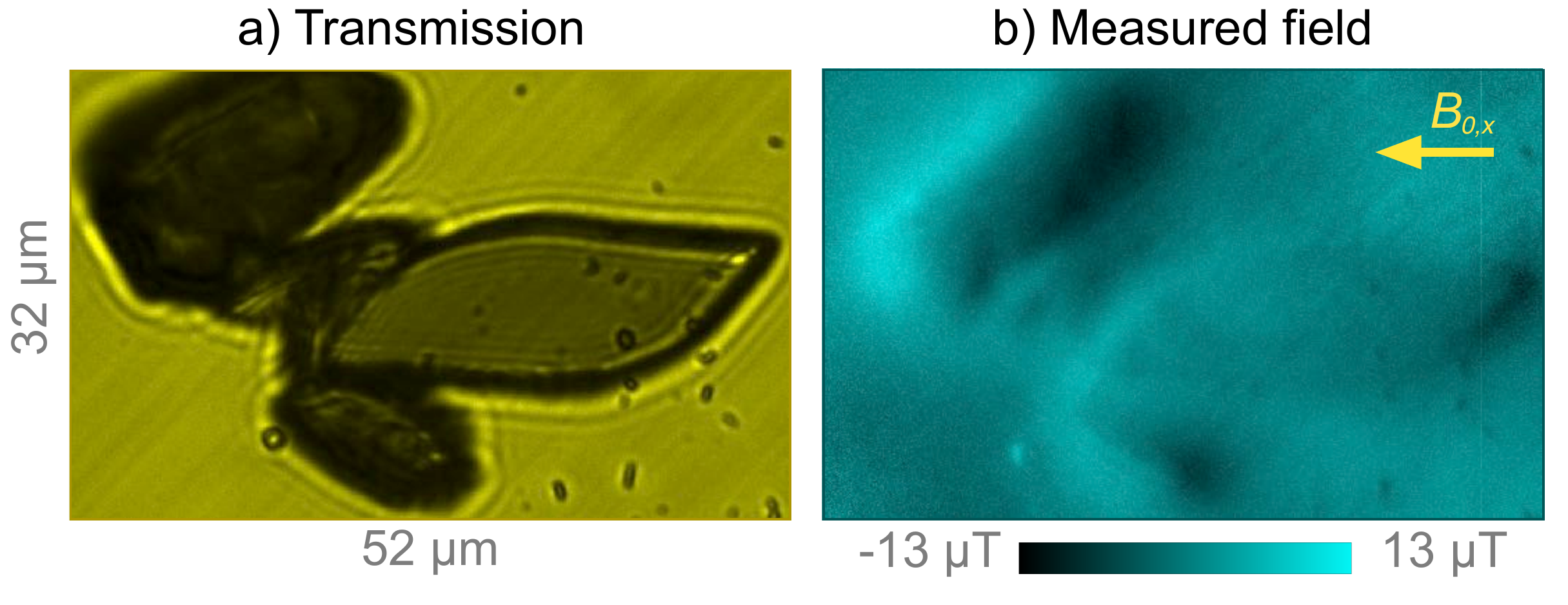}
    \end{center}
    \caption{\label{fig:hemin}
\textbf{Magnetic imaging of hemin crystals.}
a) Brightfield transmission image of hemin crystals. 
b) Corresponding diamond magnetic microscopy image ($B_{0}=186$~mT). The in-plane component of the applied field is labeled with an arrow. An out-of-plane component of similar magnitude is also present. 
}
\end{figure}

\subsection{\label{sec:hemin} Diamond magnetic microscopy of hemin}

Our diamond magnetic microscope is also capable of imaging paramagnetic substances other than hemozoin. Of possible interest are hemin microcrystals, which are used as a pharmacological agent for treating porphyria~\cite{Attarian2017}. Hemin crystals have a similar magnetic susceptibility to hemozoin but are much larger (up to 30~\textmu m). They thus serve as a convenient test sample for imaging paramagnetism.

Figure~\ref{fig:hemin} (a) shows a brightfield transmission image of hemin crystals dispersed on a diamond substrate. Figure~\ref{fig:hemin} (b) shows the corresponding diamond magnetic image. The magnetic pattern of these crystals appears different from the field patterns presented elsewhere in this manuscript. This is because a different diamond substrate (with 1.7~\textmu m thick NV layer) was used which is polished with [100] surfaces. All four NV axes in this substrate are directed $55^{\circ}$ from the surface normal, and the magnetic field was applied along one of these NV axes. This geometry leads to more asymmetric patterns than the in-plane field geometry used elsewhere. The pattern amplitudes are consistent with a magnetic susceptibility of $\chi\approx3\times10^{-4}$, similar to that of hemozoin.

\section{\label{sec:unc}Uncertainty in susceptibility}
For both natural and synthetic hemozoin, we estimate a relative uncertainty in $\chi$ of $\lesssim25\%$ due to imperfect assumptions in NV layer distribution, image blur, and hemozoin crystal dimensions/shape. The image blur is incorporating by convolving the magnetic image with a 2D Gaussian kernel. Some of these assumptions are conflated; for example, image blur and crystal height both contribute to the width of the line cuts found in Fig.~\ref{fig:zoom}(d) and Fig.~\ref{fig:zoomsyn}(d). We found the best agreement when using a blur of $0.5$~\textmu m FWHM and a crystal height of 200 nm. However we observed some magnetic features with a FWHM of $0.4$~\textmu m, e.g. n5 in Fig.~\ref{fig:zoom}(d), which is closer to the expected diffraction-limited resolution of our microscope ($\sim390~{\rm nm}$). This suggests the image blur is either spatially varying (due to imaging aberrations) or the extra modeled blur is compensating imperfections in another assumption, such as crystal height. We found that reducing the image blur to $0.4$~\textmu m lead to a best-fit estimate of $\chi=3.2\times10^{-4}$, which is $\lesssim10\%$ lower than the estimate made in the main text. However the overall agreement using those parameters was slightly worse than for the assumptions of $0.5$~\textmu m blur, $\chi=3.4\times10^{-4}$ made in the main text. By comparing the best-fit values of $\chi$ under various reasonable perturbations to the model, we arrived at an uncertainty of $\lesssim25\%$.

Another source of systematic uncertainty is that we treat $\chi$ as an isotropic quantity. Two factors suggest $\chi$ is actually anisotropic~\cite{Butykai2013}: (i) the crystals are elongated, and therefore may be more susceptible to magnetize along their long axes, and (ii) the iron centers in hematin dimers have $\rm C_{4v}$ symmetry, Fig.~\ref{fig:hems}(c), which is largely preserved after crystal formation. The latter suggest a hard axis and an easy plane whereas the former suggests all three components of magnetization may be different. In Ref. \cite{Butykai2013}, the authors made the hard axis/easy plane assumption and found a magnetic anisotropy of $M_{\rm hard-axis}/M_{\rm easy-plane}=1.16\pm0.03$ at room temperature. This variation is smaller than our estimated uncertainty due to imperfect model assumptions, but may be a contributing factor to the slight disagreement for some crystals, such as n2 [see Fig.~\ref{fig:zoom}(d)]. None of these uncertainties come close to being large enough to explain the behavior of n5 or the other presumed superparamagnetic nanoparticles.

\section{\label{sec:super}Supermagnetism}
Several points are made in the main text which support the argument that five outlier nanoparticles (including n5) are superparamagnetic. Here we outline four more pieces of evidence supporting this argument:

1. The nanoparticles are small. For two of the features we do not see anything in the SEM images at the locations of the magnetic features. The other features do have small corresponding particles in their SEM (or confocal reflectance) images, but we can't be sure they are responsible for the magnetic features. 

2. All five particles have zero coercivity, i.e. $\Delta B (0)=0$. Thus they are unlikely to be ferromagnetic.

3. The magnetic patterns of these particles are always aligned in the same way as the features from the linear paramagnetic crystals. This further decreases the likelihood these particles are ferromagnetic since they appear to be magnetized by the applied field even at fields below 0.1 T.

4. The magnetizing field (the $B_0$ field needed to magnetize the particles to half their maximum saturation magnetization) is $\sim0.1$ T. This is a reasonable value for superparamagnetic particles~\cite{Kolhatkar2013}.

\section{\label{sec:biblsup}Bibliography for Supplementary Notes}
\putbib
\end{bibunit}

\end{document}